\documentclass[reprint,aps,pra,superscriptaddress]{revtex4-2}

\usepackage{amsmath,amsthm,amssymb,amscd,float,graphicx}
\usepackage[T1]{fontenc}
\usepackage{lmodern}
\usepackage[colorlinks,linkcolor=teal,citecolor=teal,urlcolor=teal]{hyperref}
\usepackage{lipsum}
\usepackage[caption=false]{subfig} 
\usepackage{braket}
\usepackage{algorithm} 
\usepackage{algpseudocode}
\usepackage[normalem]{ulem}
\usepackage{dsfont}
\usepackage[dvipsnames]{xcolor}
\usepackage{bbm}
\usepackage{mathtools}

\newtheorem{theorem}{Theorem}[section]
\newtheorem{lemma}[theorem]{Lemma}

\newcommand{\tr}{\mathrm{tr}}
\newcommand{\eg}{\textit{e.g.},\ }
\newcommand{\ie}{\textit{i.e.},\ }
\newcommand{\cC}{{\mathcal C}}

\newcommand{\cF}{{\mathcal F}}
\newcommand{\cX}{{\mathcal X}}

\newcommand{\la}{\lambda}

\newcommand{\iu}{{\rm i}}
\newcommand{\id}{I} 

\begin{document}

\title{Simulation of noisy quantum circuits using frame representations}

\author{Janek Denzler}
\affiliation{Dahlem Center for Complex Quantum Systems, Freie Universit\"at Berlin, 14195 Berlin, Germany\looseness=-1}
\author{Jose Carrasco}
\affiliation{Dahlem Center for Complex Quantum Systems, Freie Universit\"at Berlin, 14195 Berlin, Germany\looseness=-1}
\author{Jens Eisert}
\affiliation{Dahlem Center for Complex Quantum Systems, Freie Universit\"at Berlin, 14195 Berlin, Germany\looseness=-1}
\author{Tommaso Guaita}
\email{tommaso.guaita@fu-berlin.de}
\affiliation{Dahlem Center for Complex Quantum Systems, Freie Universit\"at Berlin, 14195 Berlin, Germany\looseness=-1}

\date{\today}
\begin{abstract}
One of the core research questions in the theory of quantum computing is to find out to what precise extent the classical simulation of a noisy quantum circuits is possible and where potential quantum advantages can set in. In this work, we present a unified framework for the classical simulation of quantum circuits based on frame theory, encompassing and generalizing a broad class of existing simulation strategies. Within this framework, the computational cost of a simulation algorithm is determined by the one-norm of an associated quasi-probability distribution, providing a common quantitative measure across different simulation approaches. This enables a comprehensive perspective on common methods for the simulation of noisy circuits based on different quantum resources, such as entanglement or non-stabilizerness. It further provides a clear scheme for generating novel classical simulation algorithms. Indeed, by exploring different choices of frames within this formalism and resorting to tools of convex optimization, we are able not only to obtain new insights and improved bounds for existing methods -- such as stabilizer state simulation or Pauli back-propagation -- but also to discover a new approach with an improved performance based on a generalization of the Pauli frame. We, thereby, show that classical simulation techniques can directly benefit from a perspective -- that of frames -- that goes beyond the traditional classification of quantum resources.
\end{abstract}

\maketitle

\section{Introduction}
The classical simulation of quantum devices plays a central role in understanding the delicate boundary between classical and quantum computational power. 
In a regime where efficient classical simulation is possible, one cannot expect 
computational advantages to emerge from a quantum setting. For this reason,
 a wealth of methods have been developed to simulate quantum circuits on classical hardware, each grounded in different operational pictures of quantum mechanics and tailored to specific computational goals. A useful and highly versatile mathematical language for connecting these approaches is provided by {\em frame theory}. Frames generalize orthonormal bases by allowing redundancy, with established applications in wavelet analysis~\cite{wavelets} and quantum-state tomography, including 
\emph{symmetric informationally complete positive operator-valued measures}
(SIC-POVMs)~\cite{Renes2004,Scott2006}.

In order to substantially extend the regime of efficient classical simulation of quantum systems in a fresh direction,
in this work we present a framework, based on frame theory, that captures in a unified language several important families of existing simulation methods for noisy quantum circuits~\cite{ferrie_framed_2009,Ferrie_2011, pashayan_estimating_2015}. It provides a clear way to analyse them and ---crucially--- an approach to systematically extend them. We directly exploit this perspective by introducing two new frame-based simulations algorithms ---a product-state frame for Schrödinger-picture simulation, and an extended Pauli frame for Heisenberg-picture simulation--- which improve upon previously known methods in different noise regimes.

On the highest level, one can differentiate simulation tasks based on the answer to the fundamental question: what does it mean to simulate a quantum computation classically?
This answer may include estimating expectation values of observables, computing output probabilities (strong simulation), or sampling from the output distributions (weak simulation). 
Then, one may also broadly distinguish simulation approaches based on whether they adhere to the Schrödinger or Heisenberg picture.

In the Schrödinger picture, the focus is on constructing classical representations of the evolving quantum state. These representations ---ranging from tensor networks such as \emph{matrix product states} (MPS) and \emph{matrix product operators} (MPO)~\cite{gao_efficient_2018, vovk2024, chen_optimized_2024, sander_large-scale_2025, cichy2025classicalsimulationnoisyquantum} to linear combinations of polynomially many algebraically defined structures such as stabilizer states~\cite{bravyi_improved_2016, bravyi_simulation_2019}, Gaussian states~\cite{dias_classical_2024, cudby_gaussian_2024}, \emph{etc.}--- serve as a basis for estimating observables or reproducing output distributions. Each of these cases poses its own computational challenges and is sensitive to the structure of the quantum state.
Conversely, simulation methods based on the Heisenberg picture target the evolution of observables rather than states. These include prominent techniques such as operator back-propagation, which are particularly effective for the estimation of expectation values of Pauli strings~\cite{fontana_classical_2025, angrisani_classically_2024, angrisani_simulating_2025, shao_simulating_2024, schuster2024polynomialtimeclassicalalgorithmnoisy, gonzalez-garcia_pauli_2025}. 
Despite their substantial conceptual differences, these approaches exhibit a shared behaviour: the classical cost of simulation grows exponentially with certain quantifiers of ``quantumness'' ---be it entanglement, magic, coherence, non-Gaussianity, \emph{etc}. Simulation is therefore classically efficient only if these resources remain logarithmically bounded in system size.

The presence of noise, however, complicates this landscape. Intuitively, as noise levels increase, quantum computations lose their advantage and become more amenable to classical simulation (although in practical situations it may not always be obvious how to leverage the noise to simplify simulation). This leads to the notion of an inverse threshold: a critical error rate $p_{\rm cl}$ above which a quantum computation becomes efficiently simulable classically, regardless of the intrinsic quantum resources it employs. This concept stands in contrast to the traditional error-correction threshold $p_{\rm ec}$, below which fault-tolerant quantum computation becomes possible. Importantly, the inverse threshold is not universal ---it depends both on the noise model and on the specific type of simulation under consideration, whether strong, weak, or expectation-value-based.
We further note that to date most insights into circuit simulability have been derived in an average-case setting, involving ensembles of random circuits in certain architectures ---typically brick-wall structures, local connectivity in one dimension, \emph{etc}.

Within this classical simulation context, this work focuses on the task of computing observable expectations up to additive precision. In order to achieve this, we employ a unifying framework based on elements of frame theory and quasi-probability distributions (see ideas discussed \textit{e.g.} in references~\cite{pashayan_estimating_2015, seddon_quantifying_2021, yashin_minimal_2020, Koukoulekidis2022fasterborn, Kulikov_MinimizingNegativity}), which we here adapt explicitly to the simulation of noisy quantum circuits, both in the Schr\"odinger and in the Heisenberg picture. We find this approach offers a structured and comprehensive perspective on classical simulation across different pictures, strategies and types, by interpreting many existing methods as specific instances of a unified simulation strategy, each simply characterized by a different choice of \emph{frame} to represent operators. From this point of view, the efficiency of a given simulation algorithm can be straightforwardly analyzed by relating it to the efficiency of a Monte Carlo sampling process, constructed from the given frame representation. This observation immediately allows us to explicitly derive inverse thresholds for a variety of algorithms on worst case circuits composed from common gate-sets, without assumptions on statistical averaging over circuit ensembles.

The power of this framework however goes beyond this conceptual formalisation. Most importantly, it provides the correct perspective to derive concrete algorithmic improvements beyond known methods. 
Indeed, it
reveals that we are not bound to the traditional choices of frames (Pauli operators, stabilizer states, \emph{etc}.), but there is rather a broad ---and so far mostly unexplored--- space for developing novel simulation methods based on alternative frame choices. We demonstrate that exploring this space is a worthwhile endeavour by identifying two new frames based on product operators that provably improve on the performance of existing methods, both in the Schr\"odinger and in the Heisenberg picture.

In the rest of this paper, after reviewing some preliminary notions on classical simulation of noisy quantum circuits in Section~\ref{sec:preliminaries}, we introduce our main technical contributions, which can be summarized as follows. 

Firstly, based on the previously introduced concepts of frame theory and quasi-probabilities, in Section~\ref{sec:simulation-with-frames} we establish a conceptually unified framework specific for noisy circuit simulation. In particular:
\begin{itemize}
    \item We formulate a Monte Carlo simulation approach for noisy quantum circuits, based exclusively on the choice of a suitable frame representation in the Schrödinger or Heisenberg picture. We show that the simulation cost is governed by a single quantity, namely a one-norm that can be efficiently computed from the frame representation of the required gate-set.
    \item Within this framework, we establish the general properties that frame operators must satisfy to produce valid Schrödinger and Heisenberg picture simulation algorithms, further clarifying the different capabilities and limitations of these two approaches. This establishes a general theoretical framework within which new frame-based simulation algorithms can be developed.
\end{itemize}

Second, having established these notions we showcase their power by employing them in the analysis of several examples of simulation algorithms, based on both known and new frame choices. These results, which we divide into Schr\"odinger picture (Section~\ref{sec:results-schroedinger}) and Heisenberg picture (Section~\ref{sec:results-heisenberg}) methods, include the following:
\begin{itemize}
    \item We fully analyse previously proposed frame choices, such as the diagonal and dyadic stabilizer frames (in the Schr\"odinger picture, Section~\ref{sec:stabiliser-frames}) or the Pauli frame (in the Heisenberg picture, Section~\ref{sec:pauli-frame}), deriving the corresponding inverse thresholds. To do this we develop convex-optimization methods for computing the previously unknown optimal frame decompositions.
    \item We introduce the product-state frame for Schr\"odinger picture simulation and analyse its performance, establishing new inverse thresholds related to separability-preserving channels (Section~\ref{sec:product-frame}).
    \item We introduce a new optimal product-operator frame for Heisenberg picture simulation which generalizes the Pauli frame, leading to an improved inverse threshold under various noise models (Section~\ref{sec:extended-pauli-frame}).
\end{itemize}

These results establish that unconventional frame choices can lead to improved simulation performance, prompting further research into other potential frame definitions. In Section~\ref{sec:outlook} we conclude by discussing what other promising choices deserve more in-depth study to complete the picture outlined above.

\section{Preliminaries on classical simulation} \label{sec:preliminaries}

Among the different possible definitions of classical simulation, in this work we are specifically interested in the following setting. We consider quantum circuits composed of a polynomial number of gates chosen from a fixed gate-set, composed of gates acting on a constant number of qubits (usually one and two qubit gates). We will generally consider universal unitary gate-sets (\eg Clifford+$T$). 

On top of this, we then assume that the gates from the given gate-set may be noisy. This means that each gate is replaced by a non-unitary quantum channel only approximately close to the target unitary gate. Usually, we will construct these noisy gates by composing the unitary channel with a standard noise channel (\eg depolarizing noise) parametrised by a noise strength $p$. However one can also straightforwardly consider more general scenarios, as we shall see.

Once this noisy gate-set is fully specified, the problem is then well identified: we consider all possible circuits that can be built by concatenating $m=\mathrm{poly}(n)$ such noisy gates on $n$ qubits. Note, in particular, that we do not make any further assumptions on the circuit architecture: we do not require a specific connectivity, nor that the noise is applied uniformly to all qubits after each circuit layer, nor do we rule out the introduction of fresh qubits during the circuit.

For any such given circuit the task is to compute, with classical computational resources, the expectation value of a specific observable $O$. That is, to compute 
\begin{equation}
    \braket{O}=\tr 
    [
    \, O \,\mathcal{C}_m \circ \cdots \circ \mathcal{C}_1(\rho_0)\,
    ], \label{eq:expectation-value}
\end{equation}
where $\mathcal{C}_1$,\dots, $\mathcal{C}_m$ are the channels representing the $m$ noisy gates in the circuit and $\rho_0$ is a fixed initial state (usually the $\ket{0^n}$ state). We further assume that it is sufficient to compute this expectation value to within an additive error $\epsilon$. We say an algorithm for this task is \emph{efficient} if its runtime scales as $\mathrm{poly}(n,m,1/\epsilon)$.

Having introduced the simulation problem that we will be addressing, let us briefly comment on its relation to other notions of simulation that are commonly discussed in the literature
\cite{bravyi_simulation_2019}. First, we stress that we are interested in the simulation of expectation values to additive precision. This is different from so-called \emph{strong simulation} (\ie computing with sufficient precision the amplitudes $\braket{o_1,\dots,o_n|\psi}$, where $\ket{o_1,\dots,o_n}$ are mutual eigenvectors of some commuting observables $O_1,\dots,O_n$ and $\ket{\psi}$ is the output state vector of the circuit) and from \emph{weak simulation} (\ie sampling from the distribution $p(o_1,\dots,o_n)=|\!\braket{o_1,\dots,o_n|\psi}\!|^2$).

While weak and strong simulation are mutually inequivalent tasks, efficient weak simulation does imply efficient simulation of expectation values to additive precision. Weak simulation is also the most natural definition of what it means to simulate an actual quantum computer (which ultimately just outputs samples from a certain measurement outcome distribution)~\cite{Wigner,StabilizerPolytope}. Nonetheless, simulation of expectation values is still an interesting and non-trivial task in itself as, for instance, estimating the expectation of a single observable on an arbitrary circuit is a BQP-complete problem.

Second, we note that many relevant recent results in the field of classical simulation of (noisy) quantum circuits are statements on the \emph{average case} simulability of ensembles of random circuits~\cite{gao_efficient_2018, aharonov_polynomial-time_2023, angrisani_classically_2024, angrisani_simulating_2025, mele_noise-induced_2024, shao_simulating_2024, fontana_classical_2025}. That is, they state that if one samples a random circuit from a given ensemble, it will be with high probability a circuit that can be efficiently simulated classically. This type of results however says nothing about the small, but possibly relevant, set of circuits that are excluded by this ``high probability'' statement. In this work, in contrast, we are instead focusing on \emph{worst case} statements. That is, we want to make statements on the simulability of all circuits that can be generated by one of the noisy gate-sets introduced above.

\begin{figure*}
    \centering
    \includegraphics[width=\linewidth, trim=10mm 30mm 5mm 30mm, clip]{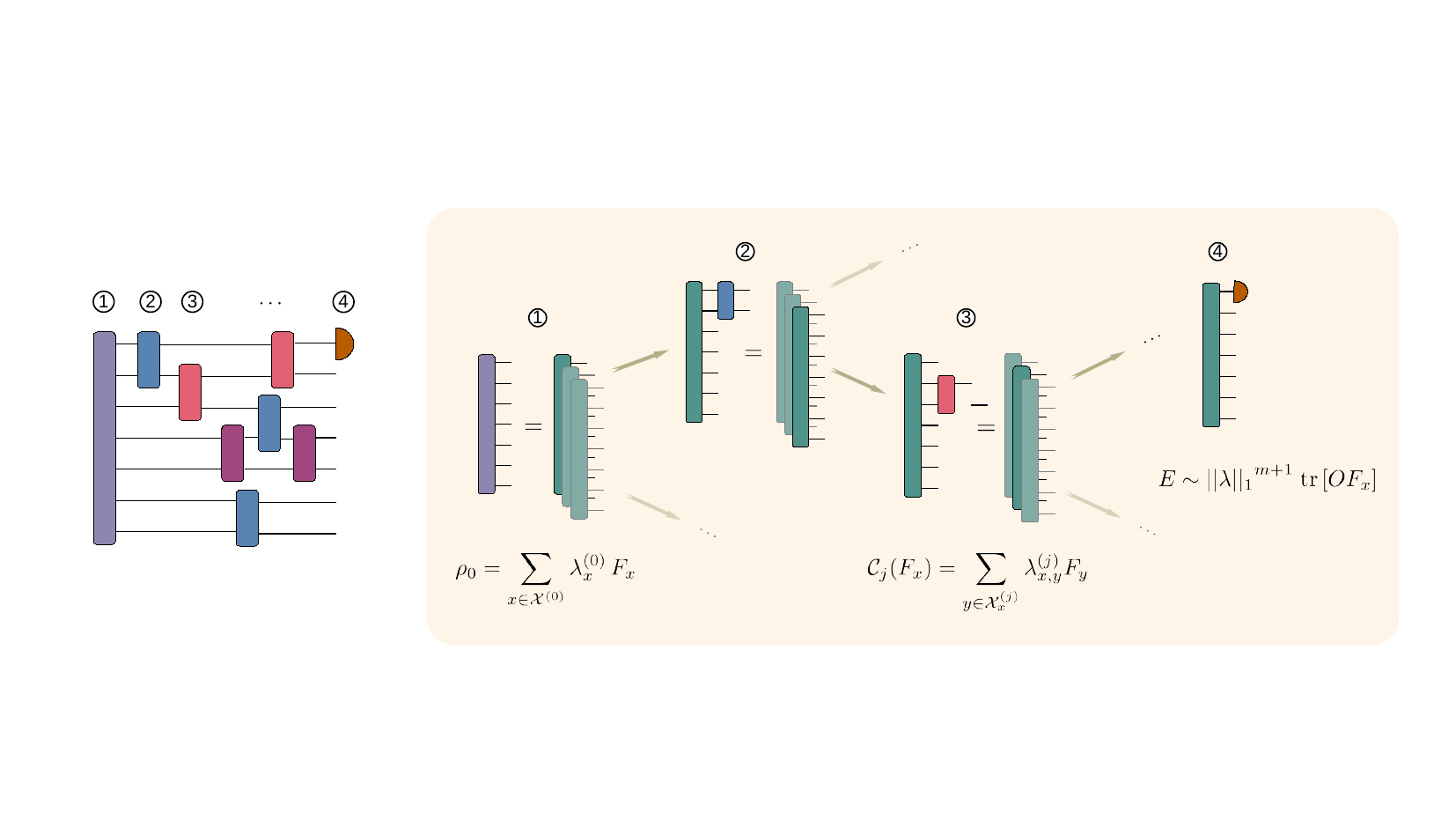}
    \caption{A schematic representation of a simulation algorithm based on the frame theory framework (represented here in the Schr\"odinger picture for illustration purposes). A path of frame elements is generated by sampling at each step a new element, based on the frame decomposition of the circuit gate acting on the previous element. At the final step, an estimator $E$ is evaluated, whose range -- and thus its sample complexity -- depends on the quantity $||\lambda||_1{}^{m+1}$.}
    \label{fig:sketch}
\end{figure*}

\section{Simulation with frames and quasi-probabilities} \label{sec:simulation-with-frames}

The main results of this work will be to introduce and analyse several methods for the classical simulation task defined above. All these methods can be collectively understood as instances of a more general simulation framework which we refer to as ``quasi-probability simulation with quantum frames''. In this section, we therefore first give a comprehensive and self-contained presentation of this framework, whose main elements have been previously introduced in Refs.~\cite{ferrie_framed_2009, pashayan_estimating_2015}. A schematic representation of the simulation algorithms in the frame theory framework can be found in Figure~\ref{fig:sketch}.

The main object that characterises any method in this framework is a so-called \emph{frame}. A frame $\mathcal{F}$ is a set of operators which spans the whole space $\mathcal{L}(\mathcal{H})$ of linear operators on the system's Hilbert space~$\mathcal H$. That is
\begin{align}
    \mathcal{F}=\left\{ F_x \,,\; \forall  x\in \cX \right\} \,,
\end{align}
where $\cX$ is an arbitrary set of indices and $F_x\in\mathcal{L}(\mathcal{H})$ are such that any linear operator $A\in\mathcal{L}(\mathcal{H})$ can be written as
\begin{align}
    A=\sum_{x\in \cX} \lambda_x^A \, F_x \label{eq:frame-expansion-operator}
\end{align}
for some set of complex coefficients $\{\lambda_x^A\}_{x\in \cX}$. 

A particular example of a frame would be a basis of $\mathcal{L}(\mathcal{H})$ such as, for instance, the Pauli basis for the space of operators on a $n$-qubit system ${\mathcal H}=({\mathbb C}^2)^{\otimes n}$. However, we do not necessarily require frames to be bases and, in particular, we do not require the frame elements $F_x$ to be all linearly independent. It follows that a frame could also be an overcomplete set of linear generators of $\mathcal{L}(\mathcal{H})$ and that the expression~\eqref{eq:frame-expansion-operator} does not uniquely identify the coefficients $\lambda_x^A$, rather there could exist many choices of $\lambda_x^A$ that represent the same operator $A$.

Given a frame, we can use it to represent not only operators in $\mathcal{L}(\mathcal{H})$ but also superoperators (\ie maps from $\mathcal{L}(\mathcal{H})$ to itself, such as quantum channels). A superoperator $\mathcal{C}$ will be represented by a ``matrix'' of coefficients $\lambda_{x,y}^{\mathcal{C}}$ depending on two frame indices such that
\begin{equation}
    \mathcal{C}(F_x)= \sum_{y\in \cX} \lambda_{x,y}^{\mathcal{C}} \, F_y \label{eq:frame-expansion-channel}
\end{equation}
for all $x\in \cX$. Again, note that this representation is in general not unique.

\subsection{Circuit evolution with frames}

The main idea of the quasi-probability method is to consider expression~\eqref{eq:expectation-value} and represent either the state or the observable as it evolves through the circuit by expanding it on the frame elements. To make this more concrete, we will for now consider the case in which we represent the evolution of the state (\ie the Schr\"odinger picture), but we will later comment on how the same concept can be applied also to representing the evolution of the observable (\ie the Heisenberg picture).

In a given frame $\mathcal{F}$, the initial state $\rho_0$ admits a decomposition $\rho_0=\sum_{x\in \cX} \lambda_x^{\rho_0} \, F_x$. 
Substituting this into equation~\eqref{eq:expectation-value} gives 
\begin{equation}
    \braket{O}= \sum_{x\in \cX} \lambda_x^{\rho_0} \: \tr \left[ O \,\mathcal{C}_m \circ \cdots \circ \mathcal{C}_1(F_x)\right]\,.
\end{equation}
We can now imagine to also have access to frame decompositions of the form~\eqref{eq:frame-expansion-channel} for all the channels $\mathcal{C}_j$ and substitute these in turn into the expression above. This leads to
\begin{equation}
    \braket{O}= \!\!\sum_{x_1,\dots,x_{m+1}}\!\! \lambda_{x_1}^{(0)}\, \lambda_{x_1,x_2}^{(1)}\cdots \lambda_{x_m,x_{m+1}}^{(m)} \: \tr \left[ O \,F_{x_{m+1}}\right]\,, \label{eq:frame-expansion-circuit}
\end{equation}
where to simplify the notation we have used the shorthand $\lambda_{x}^{(0)}\equiv \lambda_x^{\rho_0}$ and $\lambda_{x,y}^{(j)}\equiv \lambda_{x,y}^{\mathcal{C}_j}$. Note that we are here assuming that one can always find and write down these frame decompositions: we will later discuss in detail what conditions need to be satisfied such that there is actually an efficient algorithm for this, but we will for now postpone this issue to focus first on the general idea.

The expression~\eqref{eq:frame-expansion-circuit} is nothing but an equivalent rewriting of~\eqref{eq:expectation-value}. The sum over $x_1,\dots,x_{m+1}$ runs over a number of terms exponentially large in $m$, even if each $x_j$ only takes a small constant number of values. So evaluating directly the sum in~\eqref{eq:frame-expansion-circuit} does not lead to an efficient simulation of the expectation value. The idea, however, is to observe that, if we are only interested in estimating $\braket{O}$ to additive precision, then it might be sufficient to only consider a small stochastically sampled subset of terms of the sum. 

To put this Monte Carlo idea into practice, we need to identify some natural probability distribution to sample from. We do this by expressing each complex number $\lambda_{x,y}^{(j)}$ as a product of its modulus $|\lambda_{x,y}^{(j)}|$ and a complex phase $\varphi_{x,y}^{(j)}$. With this notation we can write
\begin{widetext}
\begin{align}
    \braket{O}&= \!\!\sum_{x_1,\dots,x_{m+1}}\!\! |\lambda_{x_1}^{(0)}|\,|\lambda_{x_1,x_{2}}^{(1)}| \cdots |\lambda_{x_m,x_{m+1}}^{(m)}| \,\tr \left[ O \,F_{x_{m+1}}\right]\, \varphi_{x_1}^{(0)}\varphi_{x_1,x_{2}}^{(1)}\dots\varphi_{x_m,x_{m+1}}^{(m)}\\[5mm]
    &=\!\!\sum_{x_1,\dots,x_{m+1}} \frac{|\lambda_{x_1}^{(0)}|}{\sum_{x_1'} |\lambda_{x_1'}^{(0)}|}\,\frac{|\lambda_{x_1,x_{2}}^{(1)}|}{\sum_{x_{2}'}|\lambda_{x_1,x_{2}'}^{(1)}|}  \cdots \frac{|\lambda_{x_m,x_{m+1}}^{(m)}|}{\sum_{x_{m+1}'}|\lambda_{x_m,x_{m+1}'}^{(m)}|} \; E_{x_1,\dots,x_{m+1}}
    \nonumber 
\end{align}
where in the second step we have just multiplied and divided by the normalisation terms $\sum_{x_{j+1}'}|\lambda_{x_j,x_{j+1}'}^{(j)}|$ and then collected several factors into the quantity
\begin{equation} \label{eq:E}
    E_{x_1,\dots,x_{m+1}}:= \tr \left[ O \,F_{x_{m+1}}\right]\: \varphi_{x_1}^{(0)}\varphi_{x_1,x_{2}}^{(1)}\dots\varphi_{x_m,x_{m+1}}^{(m)}\: \left(\sum_{x_1'} |\lambda_{x_1'}^{(0)}|\right) \left(\sum_{x_{2}'}|\lambda_{x_1,x_{2}'}^{(1)}|\right) \cdots \left(\sum_{x_{m+1}'}|\lambda_{x_m,x_{m+1}'}^{(m)}|\right)\,.
\end{equation}
\end{widetext}
We now observe that the following objects are well-defined probability distributions
\begin{align}
    p_0(x)&=\frac{|\lambda_{x}^{(0)}|}{\sum_{x'} |\lambda_{x'}^{(0)}|}, \label{eq:prob1}
    \\[3mm]
        p_j(x|y)&=\frac{|\lambda_{y,x}^{(j)}|}{\sum_{x'}|\lambda_{y,x'}^{(j)}|} \,,\quad \mbox{for }j=1,\dots,m,
        \label{eq:prob2}
\end{align}
namely they are positive and normalised as $\sum_x p_0(x)=\sum_x p_j(x|y)=1$. This allows us to finally write
\begin{align}
    \braket{O}&=\!\!\!\sum_{x_1,\dots,x_{m+1}} \!\!\! p_0(x_1)\cdots p_m(x_{m+1}|x_m)\, E_{x_1,\dots,x_{m+1}}\label{eq:MonteCarloEstO}\\[3mm]
    &=\mathbb{E} \, [\,E_{x_1,\dots,x_{m+1}}\,]\,,
    \nonumber 
\end{align}
where $\mathbb{E}$ denotes an expectation value with respect to the probability distribution
\begin{align}
    p(x_1,\dots,x_{m+1})=p_0(x_1)\, p_1(x_2|x_1)\cdots p_m(x_{m+1}|x_m)\,. \label{eq:distribution-product}
\end{align}

This allows us to conclude that we can evaluate the expectation value $\braket{O}$ just by drawing samples $(x_1,\dots,x_{m+1})$ from the distribution $p(x_1,\dots,x_{m+1})$ and computing the average value of the corresponding random variable $E_{x_1,\dots,x_{m+1}}$. The structure of the probability distribution~\eqref{eq:distribution-product}, which is expressed as a product of conditional probabilities, implies that we can draw each sample according to the following protocol: we first sample $x_1$ from the distribution $p_0(x_1)$, then we sample $x_2$ from the distribution $p_1(x_2|x_1)$ where $x_1$ has been fixed to the value that we sampled at the previous step, and so on for the rest of $x_3$,\dots,$x_m$.

After sampling $N$ such samples, which we label as $(x_1^{(k)},\dots,x_{m+1}^{(k)})$ for $k=1,\dots,N$, we can estimate the average value as 
\begin{align}
    \widetilde{E}=\frac{1}{N}\sum_j E_{x_1^{(k)},\dots,x_{m+1}^{(k)}}\,.
\end{align}
As one increases the number of samples $N$ this estimate will converge to the value of $\braket{O}$. The remaining question is how many such samples are needed to obtain a sufficiently precise estimate. This can be addressed by using the Hoeffding inequality~\cite{Hoeffding01031963}, which states that to obtain an estimate that is within precision $\epsilon$ with probability at least $1-\delta$, one needs to take $N$ samples with
\begin{equation}
    N=\frac{2\log(2/\delta)}{\epsilon^2} \,|E|^2 \,, \label{eq:hoeffding}
\end{equation}
where we have defined $|E|:= \sup_{x_1,\dots,x_{m+1}} |E_{x_1,\dots,x_{m+1}}|$. 

We see, therefore, that the number of required samples, and thus also the runtime of the simulation algorithm, will depend crucially on $|E|$, \ie the range of the random variable $E_{x_1,\dots,x_{m+1}}$. 
To simplify the following expressions we introduce here a one-norm notation 
\begin{align}
    ||\lambda^{(0)}||_1 &:= \sum_{x'} |\lambda_{x'}^{(0)}|, \label{eq:1norm-0} 
    \\
    ||\lambda^{(j)}_{x}||_1  &:= \sum_{x'} |\lambda_{x,x'}^{(j)}| \,,\quad \mbox{for }j=1,\dots,m \label{eq:1norm-j-x} 
\end{align}
to indicate the normalisation terms appearing in $|E|$. We can then estimate $|E|$ as
\begin{align}
    |E|&= \sup_{x_1,\dots,x_{m+1}}|\tr\left[ O F_{x_{m+1}}\right]\!| \:   ||\lambda^{(0)}||_1 ||\lambda^{(1)}_{x_1}||_1\cdots ||\lambda^{(m)}_{x_m}||_1 \nonumber \\
    &\leq {||\lambda||_1}^{m+1} \; \sup_{x}|\tr\left[ O F_{x}\right]\!|\,,\label{eq:E-range}
\end{align}
where in the second step~\eqref{eq:E-range} we have derived an upper-bound by defining the maximal one-norm value over all inputs and all gates as
\begin{align}
    ||\lambda||_1 &:=\max_j\,||\lambda^{(j)}||_1\,,\label{eq:1norm} 
\end{align}
where
\begin{align}
    ||\lambda^{(j)}||_1 &:=  \sup_x ||\lambda^{(j)}_x||_1 \,,\quad \mbox{for }j=1,\dots,m.\label{eq:1norm-j}
\end{align}

Comparing equations~\eqref{eq:E-range} and~\eqref{eq:hoeffding}, we conclude that the number of iterations of the simulation algorithm will depend on the one-norm of the coefficients $\lambda_{x,y}^{(j)}$ which represent the circuit gates in the chosen frame and in particular it will be bounded by $O(||\lambda||_1^{\,2m})$. That is, if $||\lambda||_1>1$ then the algorithm's runtime can grow exponentially in the number of gates $m$. On the other hand the runtime will remain bounded in $m$ if, for each gate in the circuit and for each input index $x$, the one-norm $||\lambda^{(j)}_{x}||_1 $ is upper bounded by $1$. 
Notice further that, if only a subset of gates in the gate-set has $||\lambda^{(j)}||_1>1$, then the cost of the algorithm scales exponentially in the number of these ``difficult to simulate'' gates (and not in the total number of gates $m$).

Finally, the following observation will be crucial for the results presented later. We have so far simply assumed that we choose a set of coefficients $\lambda_{x,y}^{(j)}$ that satisfy~\eqref{eq:frame-expansion-channel} for all the gates in the circuit. For any such choice, the analysis above is valid and the resulting algorithm will have a runtime depending on the value of $||\lambda||_1$ for this specific choice. We note, however, that for general frames the choice of coefficients is not unique, so there is a certain freedom in the algorithm design depending on this choice. It is then clear that, ideally, one should choose, among all possible decompositions of each circuit gate, the ones that give the minimal value of $||\lambda^{(j)}||_1$. While it is true that efficiently solving this optimization problem is not always possible, it is nonetheless also true that any choice that improves the value of $||\lambda^{(j)}||_1$ will correspondingly reduce the simulation cost~\cite{yashin_minimal_2020, Koukoulekidis2022fasterborn, Kulikov_MinimizingNegativity}. 
We will see in what follows that exploiting overcomplete frames to search for better decompositions can in fact be beneficial. 

This is in a way analogous to what is referred to as the ``easing'' of the sign problem in standard Monte Carlo methods \cite{Easing,SignsTerhal}. Indeed, the overall structure of the above method -- except for the use of frames -- is very similar to the way Monte Carlo methods are commonly being used in the simulation of complex quantum systems \cite{RevModPhys.73.33}. 
If expressions such as Eq.~\eqref{eq:prob1} and \eqref{eq:prob2} do not have a denominator, as $\lambda_x^{(0)}$ and
$\lambda_{y,x}^{(j)}$
are readily non-negative and normalized, then one refers to the problem as a ``sign-free'' and efficient sampling is possible
\cite{RevModPhys.73.33,PhysRevLett.94.170201}. If this is not the case, then the normalizations in Eq.~\eqref{eq:prob1} and \eqref{eq:prob2} commonly lead to an exponential sample complexity. It is known that being sign-free is a basis dependent property \cite{Biermann, Signs1,Easing,SignsTerhal}, so effort has been made to understand how to make problems sign-free by suitable basis choices or how to ease the sign problem, quantified in terms of one-norms, and what the precise computational complexity of this process is \cite{Signs1,Easing,SignsTerhal}. Here, we face a partially related task where the sampling overhead is reduced by searching for optimal representations in a fixed frame, which possibly encompasses many bases at the same time.

\subsection{Simulation in the Schr\"odinger picture}  \label{sec:simulation-sch}

In the previous discussion we have introduced the general idea of how a stochastic Monte Carlo algorithm for circuit expectation values can be derived by representing the circuit on a chosen frame. We have shown how ultimately the algorithm's complexity will depend on the value of a certain one-norm of the coefficients representing the circuit's gates in the given frame. We will now elaborate more in detail on the precise conditions that a frame must satisfy, such that the algorithms described abstractly above can be indeed instantiated efficiently in practice. 
The following conditions establish the most general theoretical framework for defining valid frame choices and developing simulation algorithms based on them. We start by considering simulation in the Schr\"odinger picture.

First, we need a frame that can be efficiently represented classically. That is, we require that each frame element $F_x$ is identified  by a label $x$ that can be expressed as a set of classical parameters of size at most $\mathrm{poly}(n)$. Note that the total number of frame elements could nonetheless be exponentially large in $n$ or, in principle, even infinite for continuous parametrizations. 

Given the frame, we need to be able to decompose the initial state and each (noisy) gate of the circuit as linear combinations of some subset of frame elements, that is,
\begin{align}
    \rho_0&=\sum_{x\in \cX^{(0)}} \lambda_x^{(0)} \, F_x \,, \label{eq:frame-expansion-state-sch} \\
    \mathcal{C}_j(F_x)&= \sum_{y\in \cX^{(j)}_x} \lambda_{x,y}^{(j)} \, F_y \label{eq:frame-expansion-gate-sch}
\end{align}
for all inputs $F_x\in\mathcal{F}$. Here $\cX^{(0)}\subseteq \cX$ and $\cX^{(j)}_x\subseteq \cX$ are subsets of indices that represent the support of each of these decompositions. A valid choice of decompositions~\eqref{eq:frame-expansion-state-sch} and~\eqref{eq:frame-expansion-gate-sch} should be identifiable efficiently classically. That is, we require that both the supports $\cX^{(0)}$ and $\cX^{(j)}_x$ and the coefficients $\lambda_x^{(0)}$ and $\lambda_{x,y}^{(j)}$ should be classically efficiently computable in terms of the inputs $x$ and $y$. 
We will further need to be able to efficiently sample elements out of these supports, according to probability distributions proportional to $|\lambda_x^{(0)}|$ and $|\lambda_{x,y}^{(j)}|$.
Various structures could lead to this property, but in most example that we treat this will be because the sets $\cX^{(0)}$ and $\cX^{(j)}_x$ only contain a constant number of elements in the system size $n$.

Finally, to evaluate the random variable $E_{x_1,\dots,x_{m+1}}$, we need to be able to compute $\tr\left[ O F_{x}\right]$ for the observable of interest $O$ and for any frame element $F_x$. This should be computable efficiently classically for any $x$ and its value should be bounded, to prevent the range~\eqref{eq:E-range} from becoming inefficiently large. That is, we require
\begin{equation}
    |\tr\left[ O F_{x}\right]|\leq \mathrm{poly}(n)\, \quad \forall x\in \cX\,. \label{eq:expectation-bound-sch}
\end{equation}

A choice of frame that satisfies all the requirements above for the given circuit will produce a stochastic algorithm for the expectation value $\braket{O}$, according to the protocol discussed in the previous subsection. In this algorithm, each sample can be produced with a classically efficient runtime of order $\mathrm{poly}(n,m)$. The total complexity of the algorithm will then be determined by the required number of these samples, which, as we have discussed, will scale as $O(\|\lambda\|_1^{\,2m})$ and thus depends on the value of $\|\lambda\|_1$ of the considered gate decompositions~\eqref{eq:frame-expansion-gate-sch}.

\subsection{Simulation in the Heisenberg picture} \label{sec:simulation-heis}

A very similar analysis can be performed when treating the problem in the Heisenberg picture, that is rewriting~\eqref{eq:expectation-value} as
\begin{equation}
    \braket{O}=\tr [\, \rho_0 (\,\mathcal{C}^*_1 \circ \cdots \circ \mathcal{C}^*_m(O) )].
\end{equation}
Here, $\mathcal{C}^*$ denotes the adjoint of the channel $\mathcal{C}$ with respect to the Hilbert-Schmidt inner-product on $\mathcal{L}(\mathcal{H})$. The simulation algorithm is constructed completely analogously to before, by first decomposing $O$ on the frame elements and then sequentially applying each gate $\mathcal{C}^*_j$ and again decomposing the resulting outcome. The final estimator for the desired expectation value will have the form~\eqref{eq:E}, where we replace $\tr \left[ O \,F_{x_{m+1}}\right]$ by $\tr \left[ \rho_0 \,F_{x_{m+1}}\right]$.

The conditions on a valid choice of frame follow analogously to the ones discussed above. We assume again to have a frame $\mathcal{F}$ where each element can be labelled efficiently classically. We now require that a choice of decompositions of the form 
\begin{align}
    O&=\sum_{x\in \cX^{(0)}} \lambda_x^{(0)} \, F_x \,, \label{eq:frame-expansion-state-heis}\\
    \mathcal{C}^*_j(F_x)&= \sum_{y\in \cX^{(j)}_x} \lambda_{x,y}^{(j)} \, F_y \,,\label{eq:frame-expansion-gate-heis}
\end{align}
can be identified efficiently classically as for all gates $\mathcal{C}^*_j$ and all inputs $F_x\in\mathcal{F}$. We further require to be able to sample efficiently from the supports $\cX^{(0)}$ and $\cX^{(j)}_x$, possibly because they all have a constant size. Finally, we require that the quantity $\tr \left[ \rho_0 \,F_{x}\right]$ can be evaluated efficiently classically for all $x\in \cX$ and it satisfies
\begin{equation}
    |\tr\left[ \rho_0 F_{x}\right]|\leq \mathrm{poly}(n)\, \quad \forall x\in \cX\,. \label{eq:bounded-expectation-heis}
\end{equation}

If all the requirements above are satisfied for the given circuit, then a stochastic algorithm for the expectation value $\braket{O}$ exists, where each sample can be produced with a classically efficient runtime of order $\mathrm{poly}(n,m)$. The total complexity of the algorithm will then be determined by the required number of these samples, which will scale as $O(\|\lambda\|_1^{\,2m})$ and thus depends on the value of $\|\lambda\|_1$ of the considered gate decompositions~\eqref{eq:frame-expansion-gate-heis}.

Note that the conditions summarized here are not completely equivalent to the ones for Schr\"odinger picture simulation. That is, a frame choice that is valid for the latter will not necessarily be suitable for Heisenberg picture simulation. The existence of these alternative choices can lead to different and complementary results, which we discuss below.

\section{Results in the Schr\"odinger picture}\label{sec:results-schroedinger}

A key feature of the general formalism introduced in the previous sections is that any definition of a frame leads to a different simulation method. Our main results are now derived by analysing in detail the algorithms that can be constructed with some specific choices of frames.  In this section, we focus on some frame choices suitable to be used in the Schr\"odinger picture, that is satisfying the requirements discussed in Section~\ref{sec:simulation-sch}. We begin with two frames based on stabilizer states introduced in Refs.~\cite{seddon_quantifying_2021, HowardCampbell2017}, developing an explicit method for applying them to circuit simulation. We then also construct a new approach based on product state frames. In all these cases, we show explicitly how to optimize $||\lambda||_1$ over frame decompositions. 

\subsection{Stabilizer frames} \label{sec:stabiliser-frames}
Let us consider frames, suitable to be used in the Schr\"odinger picture, where the frame elements $F_x$ are constructed out of stabilizer state vectors~\cite{gottesman_stabilizer_1997}. This can be done in two distinct ways.

\subsubsection{Diagonal stabilizer frame} \label{sec:stabiliser-frames-diag}
First, we have the frame $\cF=\{F_x\}_{x\in \cX}$ with frame elements of the form $F_x=\ket{\psi_x}\!\bra{\psi_x}$ for normalized stabilizer state vectors $\ket{\psi_x}$~\cite{HowardCampbell2017}. For example, for $n=1$ we have six elements $\ket0\!\bra0$, $\ket1\!\bra1$, $\ket\pm\!\bra\pm$, $\ket{\pm\iu}\!\bra{\pm\iu}$, respectively the $+1$ eigenstates of $\pm Z,\pm X,\pm Y$. In general, a stabilizer state vector $\ket{\psi_x}$ on $n$ qubits is identified by $n$ signed Pauli operators that pairwise commute. Therefore it can be labelled by a string $x$ of $O(n^2)$ bits: this provides a classically efficient parametrisation of the frame. The total size of the frame is $|\cX|=2^{O(n^2)}$~\cite{gottesman_stabilizer_1997,AaronsonGottesman2004}.

First, we notice that $\tr[OF_x]=\braket{\psi_x|O|\psi_x}$ can be efficiently computed if the observable $O$ is a Pauli operator~\cite{AaronsonGottesman2004}. Furthermore equation~\eqref{eq:expectation-bound-sch} is satisfied as 
\begin{equation}
|\braket{\psi_x|O|\psi_x}|\leq \|O\|.
\end{equation}
This makes the frame suitable for simulation in the Schr\"odinger picture.
We therefore study Eqs.~\eqref{eq:1norm-0}-\eqref{eq:1norm-j-x} which take the form
\begin{align}
    \rho_0&=\sum_{x\in \cX^{(0)}}\la^{(0)}_x\ket{\psi_x}\!\bra{\psi_x}\,,\label{eq:frame-expansion-state-diag}\\
    \cC_j(\ket{\psi_x}\!\bra{\psi_x})&=\sum_{y\in \cX_x^{(j)}}\la^{(j)}_{x,y}\ket{\psi_y}\!\bra{\psi_y}\,.\label{eq:frame-expansion-gate-diag}
\end{align}
For an initial stabilizer state such as $\rho_0=\ket{0^n}\!\bra{0^n}$, a trivial decomposition~\eqref{eq:frame-expansion-state-diag}  exists where $\cX^{(0)}$ contains a single element and $\|\la^{(0)}\|_1=1$.

Regarding the decomposition of gates, if a channel $\mathcal{C}_j$ can be written as a convex combination of $\mathrm{poly}(n)$ unitary Clifford channels then one can efficiently find a decomposition~\eqref{eq:frame-expansion-gate-diag} where $\cX^{(j)}_x$ contains polynomially many elements and $\|\lambda^{(j)}_{x}\|=1$. More specifically, consider $\mathcal{C}_j(\cdot)=\sum_k p_k C_k (\cdot) C_k^\dag$, where $C_k$ are Clifford unitaries. Then, because Cliffords map stabilizer states to stabilizer states, $\cC_j(\ket{\psi_x}\!\bra{\psi_x})$ is of the form~\eqref{eq:frame-expansion-gate-diag} where the coefficients $\lambda_{x,y}^{(j)}$ coincide with the positive normalised distribution $p_k$. It follows that this class of channels can be simulated with an efficient algorithm cost. For example, this is the case for any local Clifford gate affected by Pauli noise.

For what concerns general non-Clifford channels $\mathcal{C}_j$, we analyse here the case of single-qubit gates. Indeed, common gate-set choices require just one single-qubit non-Clifford gate, \eg the $T$-gate. Note, however, that the following analysis can be in principle generalized to the case of arbitrary local gates supported on a constant number $q$ of qubits. Without loss of generality, we consider the action of a single-qubit channel $\cC_j=\mathcal{I}^{\otimes (n-1)}\otimes\cC'_j$ acting on the last qubit of the system (channels acting on other qubits can be seen as permutations of this one, where permutations are Clifford operations).

We would now like to find decompositions of the form~\eqref{eq:frame-expansion-gate-diag} for this channel $\cC_j$ which are optimal in the sense of having the lowest possible $\|\lambda^{(j)}\|_1$, as this corresponds to the lowest simulation cost. We are not going to solve this problem by searching over the space of all possible decompositions, but we will rather restrict ourselves to a subset of decompositions that can be treated efficiently. We do this by observing that any stabilizer state vector on $n$ qubits can be efficiently written as
\begin{equation}
\ket{\psi^{(n)}}=(C^{(n-1)}\otimes \id)\ket{0^{n-2}}\ket{\psi^{(2)}}
\end{equation}
for some Clifford unitary $C^{(n-1)}$ acting on $n-1$ qubits and some $2$-qubit stabilizer state vector $\ket{\psi^{(2)}}$. With this, we find that $\cC_j(\ket{\psi_x}\!\bra{\psi_x})$ is given by 
\begin{align}
    &\cC_j(\ket{\psi_x}\!\bra{\psi_x})=\big(C^{(n-1)}\otimes \id\big)\nonumber\\
    &\times \Big[\ket{0^{n-2}}\!\bra{0^{n-2}}\otimes(\mathcal \id\otimes\cC'_j)\big(\ket{\psi^{(2)}}\!\bra{\psi^{(2)}}\big)\Big]\big(C^{(n-1)}\otimes \id\big)^\dag\label{eq:1-qubit-gate-on-2-qubit}
\end{align}
where we have exploited the single-qubit nature of the channel $\cC_j=\mathcal{I}^{\otimes (n-1)}\otimes\cC'_j$ to commute it past the Clifford unitary. Thus, any decomposition of $(\mathcal{I}\otimes\cC'_j)
(\ket{\psi^{(2)}}\!\bra{\psi^{(2)}})$ over $2$-qubit stabilizer states $\{\ket{\psi_y^{(2)}}\!\bra{\psi_y^{(2)}}\}$ will also give a decomposition of $\cC_j(\ket{\psi_x}\!\bra{\psi_x})$, by substituting it into equation~\eqref{eq:1-qubit-gate-on-2-qubit}. Furthermore, the latter decomposition will have a value of $\|\lambda_x^{(j)}\|_1$ equal to the one of the corresponding decomposition over 2-qubit stabilizer states.
We can now therefore focus on searching for 2-qubit decompositions with the lowest possible $\|\lambda_x^{(j)}\|_1$. 

It turns out that the solution of this restricted problem is actually optimal. Consider indeed an arbitrary decomposition of the form~\eqref{eq:frame-expansion-gate-diag}. Then, by comparing this to equation~\eqref{eq:1-qubit-gate-on-2-qubit}, we see that it must hold
\begin{align}
    &(\mathcal I\otimes\cC'_j)\big(\ket{\psi^{(2)}}\!\bra{\psi^{(2)}}\big)= \nonumber\\
    &\hspace{5mm}=\sum_y \lambda_{x,y}^{(j)}\:\big\|\!\braket{0^{n-2}|\big(C^{(n-1)}\otimes \id\big)^\dag|\psi_y}\!\big\|^2 \ket{\widetilde{\psi}_y^{(2)}}\!\bra{\widetilde{\psi}_y^{(2)}},
\end{align}
where we have observed that 
\begin{align}
    \ket{\widetilde{\psi}_y^{(2)}}:=\frac{\braket{0^{n-2}|\big(C^{(n-1)}\otimes \id\big)^\dag|\psi_y}}{\big\|\braket{0^{n-2}|\big(C^{(n-1)}\otimes \id\big)^\dag|\psi_y}\big\|}
\end{align}
are normalised 2-qubit stabilizer states. It therefore follows that, if a decomposition~\eqref{eq:frame-expansion-gate-diag} exists, then there also exists a 2-qubit decomposition of $(\mathcal I\!\otimes\!\cC'_j)\big(\ket{\psi^{(2)}}\!\bra{\psi^{(2)}}\big)$ with one-norm upper-bounded by 
\begin{equation}
\sum_y|\lambda_{x,y}^{(j)}| \,\big\|\!\braket{0^{n-2}|\big(C^{(n-1)}\otimes \id\big)^\dag|\psi_y}\!\big\|^2\leq \sum_y|\lambda_{x,y}^{(j)}|.
\end{equation}
In conclusion, the result that we obtain by searching over 2-qubit decompositions cannot be worse than what one would find by searching over all possible decompositions.

More generally, for channels acting on a constant number of qubits $q$, with arguments analogous to the one above, one can restrict the problem to optimizing over stabilizer states of $q+1$ qubits. If $q$ is constant, this can be written as an efficiently solvable optimization problem, as we show in the following Section~\ref{sec:optimization-problem} for the case $q=1$.

\subsubsection{Dyadic stabilizer frame}
The previously discussed diagonal stabilizer frame is a subset of the more general dyadic stabilizer frame~\cite{seddon_quantifying_2021}. This is the frame $\cF=\{F_{{(x_1,x_2)}}\}_{(x_1,x_2)\in \cX}$ with frame elements of the form $F_{(x_1,x_2)}=\ket{\psi_{x_1}}\!\bra{\psi_{x_2}}$ for a pair of normalized stabilizer state vectors $\ket{\psi_{x_1}}$ and $\ket{\psi_{x_2}}$. This frame also clearly admits a classically efficient parametrisation in terms of the strings $x=(x_1,x_2)$ composed of $O(n^2)$ bits. For $n=1$ we have, for example, thirty six frame elements.

As before, note that $\tr[OF_x]=\braket{\psi_{x_2}|O|\psi_{x_1}}$ can be efficiently computed if the observable $O$ is a Pauli operator~\cite{AaronsonGottesman2004,bravyi_simulation_2019}. In addition, Eq.~\eqref{eq:expectation-bound-sch} is satisfied as $|\braket{\psi_{x_2}|O|\psi_{x_1}}|\leq \|O\|$. Consider now Eqs.~\eqref{eq:1norm-0}-\eqref{eq:1norm-j-x} which take the form
\begin{align}
    \rho_0&=\sum_{x\in \cX^{(0)}}\la^{(0)}_x \ket{\psi_{x_1}}\!\bra{\psi_{x_2}}\,,\label{eq:frame-expansion-state-dyadic}\\
    \cC_j(\ket{\psi_{x_1}}\!\bra{\psi_{x_2}})&=\sum_{y\in \cX_x^{(j)}}\la^{(j)}_{x,y}\ket{\psi_{y_1}}\!\bra{\psi_{y_2}}\,.\label{eq:frame-expansion-gate-dyadic}
\end{align}
For an initial stabilizer state such as $\rho_0=\ket{0^n}\!\bra{0^n}$, a trivial decomposition exists where $\cX^{(0)}$ contains a single element and $\|\la^{(0)}\|_1=1$.

Regarding the decomposition of gates, the discussion is analogous to the diagonal case. Since Clifford unitaries map stabilizer states to stabilizer states, they also map dyadic frame elements to dyadic frame elements. Therefore, if a channel $\mathcal{C}_j$ can be written as a convex combination of $\mathrm{poly}(n)$ unitary Clifford channels then one can efficiently find a decomposition~\eqref{eq:frame-expansion-gate-dyadic} where $\cX^{(j)}_x$ contains polynomially many elements and $\|\lambda^{(j)}_{x}\|=1$ by the same argument as before.

As discussed previously, for general non-Clifford channels $\mathcal C_j$ one can, without loss of generality, consider the action of a single-qubit channel $\cC_j=\mathcal{I}^{\otimes (n-1)}\otimes\cC'_j$ on the last qubit of the system. We are interested in finding decompositions of the form~\eqref{eq:frame-expansion-gate-dyadic} for such a channel $\cC_j$ having the lowest possible $\|\lambda^{(j)}\|_1$. Similarly to Eq.~\eqref{eq:1-qubit-gate-on-2-qubit} we can write $\cC_j(\ket{\psi_{x_1}}\!\bra{\psi_{x_2}})$ as
\begin{align}
    &\cC_j(\ket{\psi_{x_1}}\!\bra{\psi_{x_2}})=\big(C_1^{(n-1)}\otimes1\big)\nonumber\\
    &\times\Big[\ket{0^{n-2}}\!\bra{0^{n-2}}\otimes(\mathcal I\otimes\cC'_j)\big(\ket{\psi_1^{(2)}}\!\bra{\psi_2^{(2)}}\big)\Big]\big(C_2^{(n-1)}\otimes\id\big)^\dag\label{eq:1-qubit-gate-on-2-qubit-dyadic}
\end{align}
where $\ket{\psi_1^{(2)}}$ and $\ket{\psi_2^{(2)}}$ are $2$-qubit stabilizer state vectors, and $C_1^{(n-1)}$ and $C_2^{(n-1)}$ are Clifford unitaries acting on $n-1$ qubits. Thus, any decomposition of $\cC'_j(\ket{\psi_1^{(2)}}\bra{\psi_2^{(2)}})$ over $2$-qubit dyadic stabilizer elements $\{\ket{\psi_{y_1}^{(2)}}\bra{\psi_{y_2}^{(2)}}\}$ will also give a decomposition of $\cC_j(\ket{\psi_{x_1}}\bra{\psi_{x_2}})$ with the same value of $\|\lambda_x^{(j)}\|_1$, by substituting it into equation~\eqref{eq:1-qubit-gate-on-2-qubit-dyadic}. Similarly, for channels acting on a constant number of qubits $q$, one can restrict the optimization to dyadic frame elements of $q+1$ qubits. As before, solutions to this restricted problem turn out to be optimal also for the generic one. In summary, the optimization problem can be efficiently solved also in this case, as we discuss more in detail in the next section.

\subsubsection{Optimal decompositions for stabilizer frames} \label{sec:optimization-problem}
Let us now analyse in more detail the concrete results that can be obtained for the simulation of noisy gate-sets with the stabilizer frames introduced above. We focus here on a standard gate-set composed of arbitrary unitary Clifford gates supplemented by the magic gate $T=\sqrt[4]{Z}$. We further model the presence of noise in the system by assuming that each such gate is followed by single-qubit Pauli noise channels acting on the same qubits as the gate.

As argued in the previous sections, the Clifford gates are efficiently simulable in both the diagonal and dyadic stabilizer frames and, assuming Pauli noise, this remains true also in the noisy case. We can thus concentrate on examining the cost of simulating the noisy $T$-gates. As a concrete example, we consider then the channel $\cC_{T, {\rm  depol.}}(\cdot)=\mathcal{N}^{\rm  depol.}_p(T (\cdot) T^\dag)$, given by a $T$-gate followed by depolarising noise of strength $p$, where 
\begin{align}
    \mathcal{N}^{\rm  depol.}_p(\rho):=(1-3p)\rho + p\, (X\rho X+\!Y\rho Y+\!Z\rho Z)\,. \label{eq:depol}
\end{align}

As discussed in the previous sections, this means that we need to numerically find optimal decompositions of the 2-qubit operator
\begin{equation}
 (\mathcal{I}\otimes \mathcal{N}^{\rm  depol.}_p) (\id \otimes T \ket{\psi_1^{(2)}}\!\bra{\psi_2^{(2)}} \id \otimes T^{\dag}),
 \label{eq:2-qubit-stabilizer-decomposition}
\end{equation} 
for arbitrary 2-qubit stabilizer state vectors $\ket{\psi_1^{(2)}}$, $\ket{\psi_2^{(2)}}$ (for the diagonal frame it is sufficient to consider $\ket{\psi_1^{(2)}}=\ket{\psi_2^{(2)}}$).
For the diagonal stabilizer frame, which is Hermitian, finding a stabilizer decomposition with minimal $||\lambda||_1$ is the well-known basis pursuit problem, which can be phrased as a linear program. For the more general dyadic stabilizer frame, which allows for complex coefficients, the optimization can be phrased as a second-order cone problem.

In particular, let $y \in \mathbb{C}^{16}$ be the vector representation of the operator \eqref{eq:2-qubit-stabilizer-decomposition} in the 16-dimensional Pauli basis of two qubit operators. Furthermore, consider the matrix $D\in\mathbb{C}^{16\times|\cX^{(2)}|}$ whose columns are the vector representations, in the same Pauli basis, of all 2-qubit frame elements (that is of all objects $\ket{\psi_x^{(2)}}\! \bra{\psi_x^{(2)}}$ in the diagonal case and $\ket{\psi_{x_1}^{(2)}}\! \bra{\psi_{x_2}^{(2)}}$ in the dyadic case). Here, $|\cX^{(2)}|$ is the dimension of this 2-qubit frame (\ie 60 in the diagonal case and 3600 in the dyadic case). Then, for every possible value of $y$ (that is, for the value of expression~\eqref{eq:2-qubit-stabilizer-decomposition} corresponding to every choice $\ket{\psi_1^{(2)}}\!\bra{\psi_2^{(2)}}\in\cX^{(2)}$), we need to solve the optimization problem
\begin{equation}
\begin{aligned}
\min_{\lambda \in \mathbb{C}^{|\cX^{(2)}|}} \quad & ||\lambda||_1 \\
\textrm{s.t.} \quad & D \lambda = y.
\end{aligned}
\tag{$\mathbf{P1}$}\label{P1}
\end{equation}

One can show that the problem $\mathbf{P1}$ can be reformulated as the equivalent problem $\mathbf{P2}$ given by
\begin{equation}
\begin{aligned}
\min_{x \in \mathbb{R}^{3d}} \quad  & \,
\begin{pmatrix} \mathbf{1} &\mathbf{0} & \mathbf{0}\end{pmatrix} x \\
\textrm{s.t.} \quad & \begin{pmatrix} \mathbf{0} & & \\ & \operatorname{Re}(D) &-\operatorname{Im}(D)  \\ & \operatorname{Im}(D) & \hspace{1em}\operatorname{Re}(D) \end{pmatrix} x = \begin{pmatrix}  \mathbf{0} \\ \operatorname{Re}(y) \\ \operatorname{Im}(y) \end{pmatrix}\\[0.5em]
& \hspace{2mm} ||(\pi_{d+i} + \pi_{2d+i})x||_2 \leq e_i^{T} x\\
&\hspace{31mm} i = 1,\dots, d,
\end{aligned}
\tag{$\mathbf{P2}$}\label{P2}
\end{equation}
where $\pi_i = e_i e_i^{T}$, $d=|\cX^{(2)}|$ and $\|.\|_2$ is the Euclidean vector 2-norm. This is now a canonical formulation of a second-order cone program which can be solved with standard solver implementations~\cite{cvxpy}, efficiently in 
$|\cX^{(2)}|$.

Furthermore, note that the problem  $\mathbf{P2}$ can also be reformulated in a Lagrange dual form \cite{Boyd}, where the minimization is replaced by a maximization. This can be helpful both to find analytical solutions for these problems, and to derive rigorous proofs of optimality for numerical approaches. We discuss dual formulations more in detail in Appendix~\ref{app:duals} and use them to derive some analytical results in Appendix~\ref{app:analytics}.

\begin{figure}
    \centering
    \includegraphics[width=\linewidth]{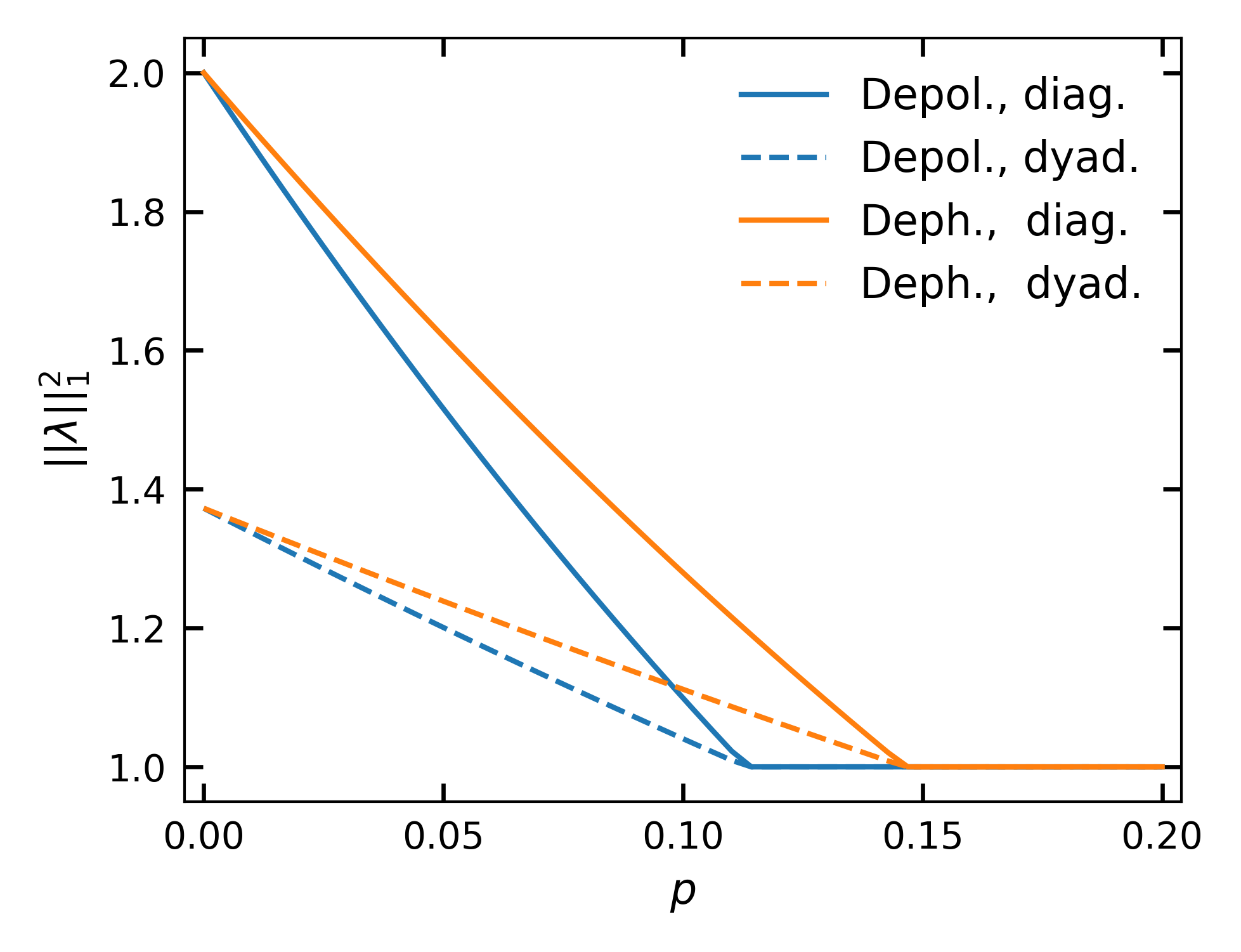}
    \caption{\textbf{Stabilizer frames.} $||\lambda^{(j)}||_1^{\,2}$ for optimal decompositions in the diagonal and dyadic stabilizer frames of the noisy $T$-gate at different depolarizing and dephasing strengths $p$. The total algorithm runtime will scale as $O(||\lambda||_1^{\,2t})$, where $t$ is the number of noisy $T$ gates in the circuit. }
    \label{fig:stabs_depol}
\end{figure}

Figure \ref{fig:stabs_depol} shows the results of these optimization problems. We plot (solid blue line) the maximal value of $\|\lambda_x^{T, {\rm  depol.}}\|_1$ over all input stabilizer frame elements for the channel $\cC_{T, {\rm  depol.}}$ as a function of the noise strength $p$. First of all, we notice that an inverse threshold behaviour is visible. That is, the noisy $T$-gate can be simulated efficiently in these frames if the noise strength exceeds a certain critical value $p_{\rm cl} \approx 0.11$, above which $\|\lambda\|_1=1$. In Appendix~\ref{app:analytics-stab}, we show that the analytical value of this threshold is $p_{\rm cl}=(6-2\sqrt{2})/28\approx 0.11$, based on known decompositions of the depolarized $T$-gate as a convex combination of Clifford channels~\cite{buhrman_new_2006}. 

Next, we observe that using the more generic dyadic frame (dashed blue line) gives a significant improvement in the simulation cost at low depolarizing strength $p$ over the diagonal frame. Nonetheless, even with this improved frame, the threshold value remains fixed at the same point. 
We will explain this last fact in Appendix~\ref{app:nogo}, showing that theorems based on magic-state distillation imply that a lower inverse threshold would be impossible for any stabilizer frame.

Finally, we can compare these results to the performance of state-of-the-art methods for the simulation of noiseless Clifford+$T$ circuits, known as \emph{stabilizer extent} methods~\cite{bravyi_improved_2016,bravyi_simulation_2019}. These methods can be naively extended to the simulation of circuits with Pauli noise by viewing the noisy circuit as an average over pure state trajectories, each containing additional Pauli operators. The best known algorithms of this type scale as $1.17^t$, where $t$ is the number of (noisy) $T$-gates in the circuit. We see that, at low noise rates, this represents a quadratic improvement over our results with the stabilizer frames. This can be traced back to some technical tools used in Ref.~\cite{bravyi_improved_2016} which strongly rely on the pure state nature of the problem (which in particular allows to compute $\braket{\psi|O|\psi}$ in a time proportional to $\ket{\psi}$'s stabilizer extent $\xi$, rather than the naively expected $\xi^2$). On the other hand, the pure state nature of these methods also implies that they cannot benefit from the presence of noise, meaning that at larger noise strengths our methods become optimal. 

We conclude by noting that the same analysis can be repeated for arbitrary Pauli noise. As an example, in Figure \ref{fig:stabs_depol} we also plot (orange lines) the case of dephasing noise
\begin{align}
    \mathcal{N}^{\rm  deph.}_p(\rho):=(1-p)\rho + p\, Z\rho Z\,,\label{eq:deph}
\end{align}
corresponding to the noisy channel $\cC_{T, {\rm  deph.}}(\cdot)=\mathcal{N}^{\rm  deph.}_p(T (\cdot) T^\dag)$.

\subsection{Product state frames} \label{sec:product-frame}
In addition to the stabilizer-based frames described in the previous section, alternative frames suitable for Schrödinger picture simulation can be constructed from other classes of efficiently classically representable states. We now discuss the example of product states. As in the case of stabilizer states, frames can be constructed in a diagonal or in a dyadic way.

\subsubsection{Diagonal product state frame}

The diagonal product state frame is defined as $\mathcal{F} = \{F_x\}$ with
\begin{align}
    F_x=\bigotimes_{k=1}^n \ket{\psi_x^{(k)}}\!\bra{\psi_x^{(k)}}.
\end{align}
Here, the tensor product runs over all qubits and each local factor $\ket{\psi_x^{(k)}}\!\bra{\psi_x^{(k)}}$ is a normalized single-qubit state.
Every product state in the frame can be specified using $\mathcal{O}(n)$ real parameters $x$. Additionally, for any Pauli or product operator $O$, the expectation $\tr{(O\bigotimes_k \ket{\psi_x^{(k)}}\!\bra{\psi_x^{(k)}})}$ can be efficiently computed and the bound~\eqref{eq:expectation-bound-sch} is satisfied, since $|\tr{(O\bigotimes_k \ket{\psi_x^{(k)}}\!\bra{\psi_x^{(k)}})}| \leq ||O||$. This makes the frame suitable for Schr\"odinger picture simulation.

In this setting, one may consider either the continuous frame that includes all product states, or choose instead a subset of finite cardinality $|\cX|$. In what follows we will discuss both the continuous case and a discrete approximation, formed by selecting a finite set of uniformly random single-qubit states from which to pick the local factors $\ket{\psi_x^{(k)}}$. The discretization enables a tractable numerical treatment in general, while for the continuous frame we have to rely on case-by-case analytical decompositions.

For an initial product state such as $\rho_0=\ket{0^n}\!\bra{0^n}$, a trivial decomposition in the frame exists containing a single element and with $\|\la^{(0)}\|_1=1$ (provided that the state vector $\ket{0}$ is included in the finite set of single-qubit states in the discrete case).
Regarding the decomposition of channels, we observe that separable channels, \ie channels whose Kraus operators factorize as $K^{(i)} = \bigotimes_{k} K^{(i)}_k$, map product states to convex combinations of product states. Consequently, any separable channel with constant Kraus rank can be efficiently simulated in the full product state frame (and we expect a similar behaviour also for suitable discrete approximations). Entangling gates, on the other hand, are not necessarily efficient to simulate with this frame. 

In general, consider a channel $\cC_j$ acting on a constant number $q$ of qubits (in the examples that follow this will be at most $q=2$). If, for instance, it acts non-trivially only on the first $q$ qubits, we can write it as $\cC_j=\cC'_j\otimes \mathcal{I}^{n-q}$ (for gates acting on other subsets of qubits the discussion follows analogously). In this setting it is clear that one can restrict the search for gate decompositions to the $q$ tensor factors where the gate has acted. That is, equation~\eqref{eq:frame-expansion-gate-sch} reduces to 
\begin{align}
    &\mathcal{C}_j'\left(\bigotimes_{k=1}^q\ket{\psi^{(k)}_x}\!\bra{\psi^{(k)}_x}\right) =  \sum_{y\in \cX^{(j)}_x} \lambda^{(j)}_{x,y} \,\bigotimes_{k=1}^q\ket{\psi^{(k)}_{y}}\!\bra{\psi^{(k)}_{y}}. \label{eq:frame-expansion-gate-prod}
\end{align}
In the case of the continuous frame, we can attempt to find analytical solutions to this equation involving a constant sized subset $\cX^{(j)}_x$ of product states. For the discrete frame, on the other hand, one can systematically solve the problem in an efficient way. Note indeed that one can without loss of generality replace $\cX^{(j)}_x$ by the set $\mathcal{X}^{(q)}$ of parametrizations of all $q$-qubit product states constructed from our finite set of normalized single-qubit states. Now equation~\eqref{eq:frame-expansion-gate-prod} relates operators on a system of constant size $q$ and similarly the size of the set $\mathcal{X}^{(q)}$ is constant.
This means that it is possible to efficiently characterize the set of coefficients $\lambda^{(j)}_{x,y}$ and find the choice that minimizes $||\lambda^{(j)}||_1$.

We can indeed proceed as we did in Section~\ref{sec:optimization-problem} for stabilizer frames.
Let $y \in \mathbb{C}^{4^q}$ be the vector representation of the left-hand side of equation~\eqref{eq:frame-expansion-gate-prod} in the Pauli basis of $q$-qubit operators. Furthermore, consider the matrix $D\in\mathbb{C}^{4^q\times|\cX^{(q)}|}$, whose columns are the vector representations in the same Pauli basis of all $q$-qubit discrete product state frame elements. Then, for all the values of $y$, corresponding to the left-hand side of~\eqref{eq:frame-expansion-gate-prod} for each choice of $\bigotimes_{k=1}^q\ket{\psi^{(k)}_x}\!\bra{\psi^{(k)}_x}$ in the discrete frame, we need to solve the optimization 
problem
\begin{equation}
\begin{aligned}
\min_{\lambda \in \mathbb{C}^{|\cX^{(q)}|}} \quad & ||\lambda||_1 \\
\textrm{s.t.} \quad & D \lambda = y.
\end{aligned}
\end{equation}
As discussed previously, this can be mapped to the second-order cone program \ref{P2} and solved efficiently.

As a concrete example, we consider a universal gate-set composed of the single-qubit Hadamard ($H$) and magic ($T$) gates, as well as the two-qubit controlled-NOT ($CNOT$) gate, where each gate is followed by a depolarizing noise channel~\eqref{eq:depol} on all qubits that it acts on. From the point of view of the continuous frame it is sufficient to focus on the $CNOT$ gate, as the single qubit ones can always be efficiently simulated. In Appendix~\ref{app:analytics-prod} we show that, for any product frame element on which the noisy gate acts, we can provide an analytical decomposition of the form~\eqref{eq:frame-expansion-gate-prod} involving $8$ frame elements. For depolarizing strength $p$, this decomposition has a value $||\lambda^{(CNOT,\rm{depol.})}||_1=3(1-4p)^2$ and is the optimal choice for the given frame.

\begin{figure}
    \centering
    \includegraphics[width=\linewidth]{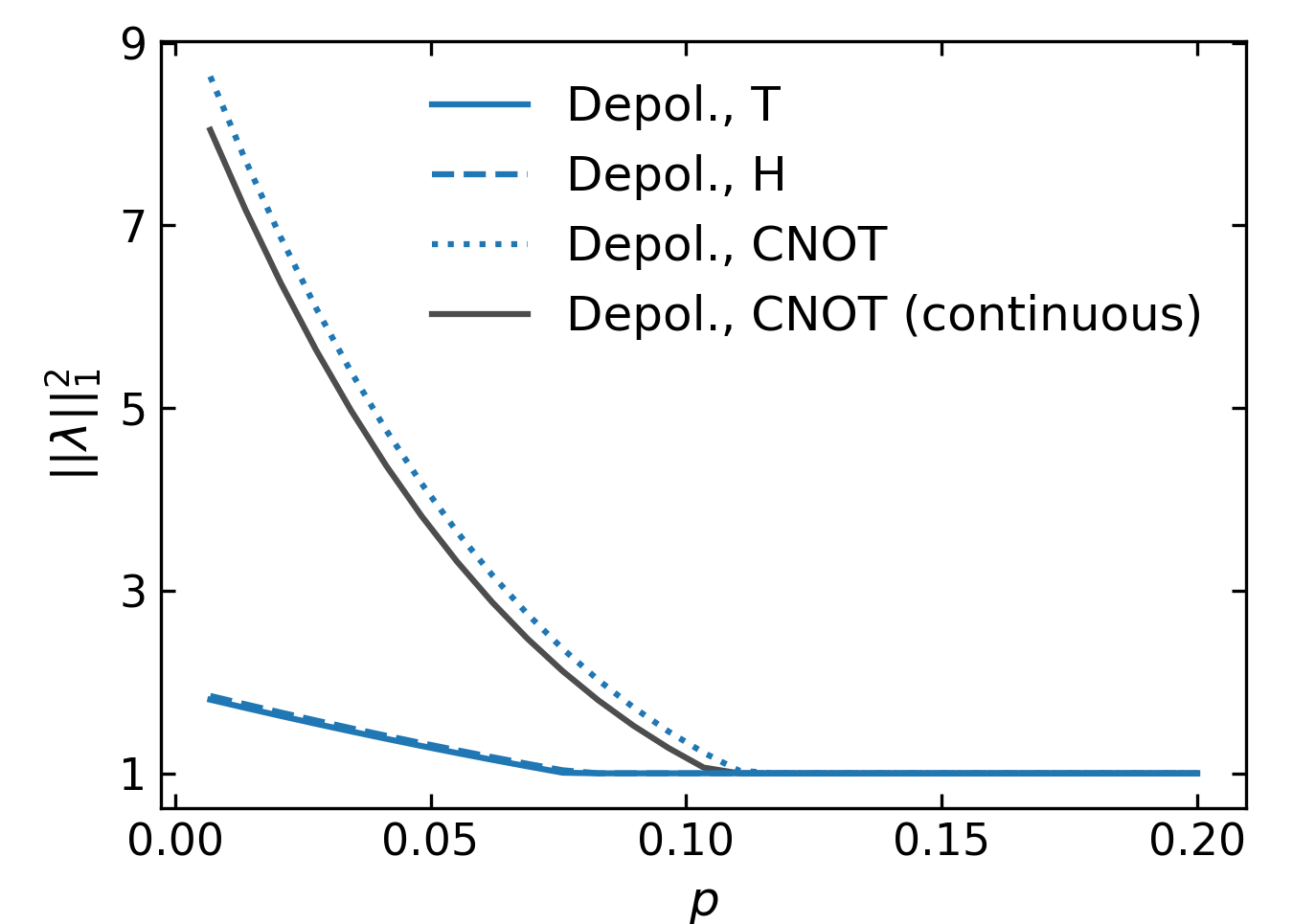}
    \caption{\textbf{Diagonal product state frame.} $||\lambda^{(j)}||_1^{\,2}$ for optimal decompositions with respect to a discrete approximation of the continuous product state frame (blue lines), plotted as a function of the depolarizing strength $p$ for the three gates in the gate-set ($H$, $T$, and $CNOT$).
    The frame is constructed from local factors selected from a finite set of $30$ uniformly random single-qubit states.
    In grey we also plot the analytical value for the noisy $CNOT$ gate derived in Appendix~\ref{app:analytics-prod} for the continuous frame of all product states.}
    \label{fig:prod_depol}
\end{figure}

For this gate-set, we also consider a numerical analysis based on the choice of a discrete set of local single-qubit states of size $30$, \ie $|\cX^{(2)}| = 30^2$. The results of the corresponding optimization are shown in Figure \ref{fig:prod_depol}. 
The maximal value of $||\lambda_x^{(j)}||_1$ across all inputs $x\in\cX$ is plotted as a function of the depolarizing strength $p$ for all channels in the gate-set.  The results show good agreement with the analytical value discussed above. The fact that the single-qubit gates $H$ and 
$T$ exhibit values $||\lambda^{(j)}||_1 > 1$ for small $p$ is due to the discrete approximation of the continuous product state manifold.

Consistent with the behaviour observed for stabilizer-based frames in Section \ref{sec:stabiliser-frames}, both methods show the existence of an inverse threshold at approximately $p_{\rm cl} \approx 0.11$. Above this threshold, the condition $||\lambda^{(j)}||_1 = 1$ holds, and all circuits composed of these gates can be efficiently simulated. The analytical decomposition, in particular, gives the value $p_{\rm cl} =  \frac{1}{4}(1-\frac{1}{\sqrt{3}})$ for the inverse threshold.

Note that, although the $CNOT$ gate is maximally entangling, its noisy counterpart $\mathcal{C}_{CNOT,{\rm  depol.}}(\cdot) = (\mathcal{N}^{\rm  depol.}_p \otimes \mathcal{N}^{\rm  depol.}_p)[CNOT (\cdot) CNOT^\dag]$ is known to become separable once the depolarizing strength $p$ is sufficiently large. As shown in Ref.~\cite{cnot_separability}, this occurs for $p\geq (8-\sqrt{8})/28\approx 0.185$. In this regime, the gate-set can be trivially simulated. 
For smaller noise strengths $p \lesssim  0.185$, the channel $\mathcal{C}_{\text{CNOT},{\rm  depol.}}$ is not separable. What our results show is that for $0.11\lesssim p \lesssim 0.185$ it is nevertheless separability-preserving -- that is, it maps all input product states to convex combinations of product states, leading to $||\lambda^{(j)}||_1 = 1$ and hence to efficient simulability.

\subsubsection{Dyadic product frame}

Analogously to the stablizer case, the dyadic product state frame is defined as $\cF=\{F_{{(x_1,x_2)}}\}_{(x_1,x_2)\in \cX}$ with frame elements 
\begin{align}
    F_{(x_1,x_2)}=\bigotimes_{k=1}^n \ket{\psi_{x_1}^{(k)}}\!\bra{\psi_{x_2}^{(k)}}.
\end{align}
Here $\ket{\psi_{x_1}^{(k)}}$ and $\ket{\psi_{x_2}^{(k)}}$ are normalized single-qubit state, for each qubit $k$. Every frame element of this type clearly admits a classical parametrisation in terms of the string $x=(x_1,x_2)$ composed of $O(n^2)$ real parameters.

As for the diagonal frame, for any product operator $O$, the expectation $\tr{(O\bigotimes_k \ket{\psi_{x_1}^{(k)}}\!\bra{\psi_{x_2}^{(k)}})}$ can be efficiently computed and its modulus is upper-bounded by $||O||$, satisfying the bound~\eqref{eq:expectation-bound-sch} and giving a frame suitable for Schr\"odinger picture simulation. Furthermore, for initial product states such as $\rho_0=\ket{0^n}\!\bra{0^n}$, a trivial decomposition in the frame exists containing a single element and with $\|\la^{(0)}\|_1=1$. Separable channels with constant Kraus rank, and in particular single qubit channels, map frame elements to convex combinations of frame elements and can thus always be efficiently simulated.

We focus therefore on entangling $q$-qubit gates. For the channel $\cC_j=\cC'_j\otimes \mathcal{I}^{n-q}$, we need to find decompositions of the form
\begin{align}
    &\mathcal{C}_j'\left(\bigotimes_{k=1}^q\ket{\psi^{(k)}_{x_1}}\!\bra{\psi^{(k)}_{x_2}}\right) =  \sum_{y\in \cX^{(j)}_x} \lambda^{(j)}_{x,y} \,\bigotimes_{k=1}^q\ket{\psi^{(k)}_{y_1}}\!\bra{\psi^{(k)}_{y_2}}. \label{eq:frame-expansion-gate-prod-dyad}
\end{align}
In principle, as for the diagonal frame, we can introduce a discretization of the frame and numerically solve the problem of finding such a decomposition. However, for two-qubit gates, the number of relevant discrete frame elements will be larger by a power of two compared to the diagonal case. This makes the numerical approach rather cumbersome even for moderately sized discretizations. For this reason, we will rather focus here only on some analytical results that can be derived in the specific case of depolarizing noise.

Indeed, if we consider the channel $\cC_{CNOT,\mathrm{depol.}}$ given by the controlled-NOT gate followed by the depolarizing noise channel~\eqref{eq:depol} on both qubits that it acts on, we find that for any dyadic input frame element $\bigotimes_{k=1}^q\ket{\psi^{(k)}_{x_1}}\!\bra{\psi^{(k)}_{x_2}}$, a decomposition of the form~\eqref{eq:frame-expansion-gate-prod-dyad} exists involving $8$ dyadic frame elements. We give this decomposition explicitly in Appendix~\ref{app:analytics_prod_dyad}, where we also prove that it achieves the optimal one-norm value for this frame, given by $||\lambda^{(CNOT,\rm{depol.})}||_1=(1+3(1-4p)^2)/2$.

\begin{figure}
    \centering
    \includegraphics[width=\linewidth]{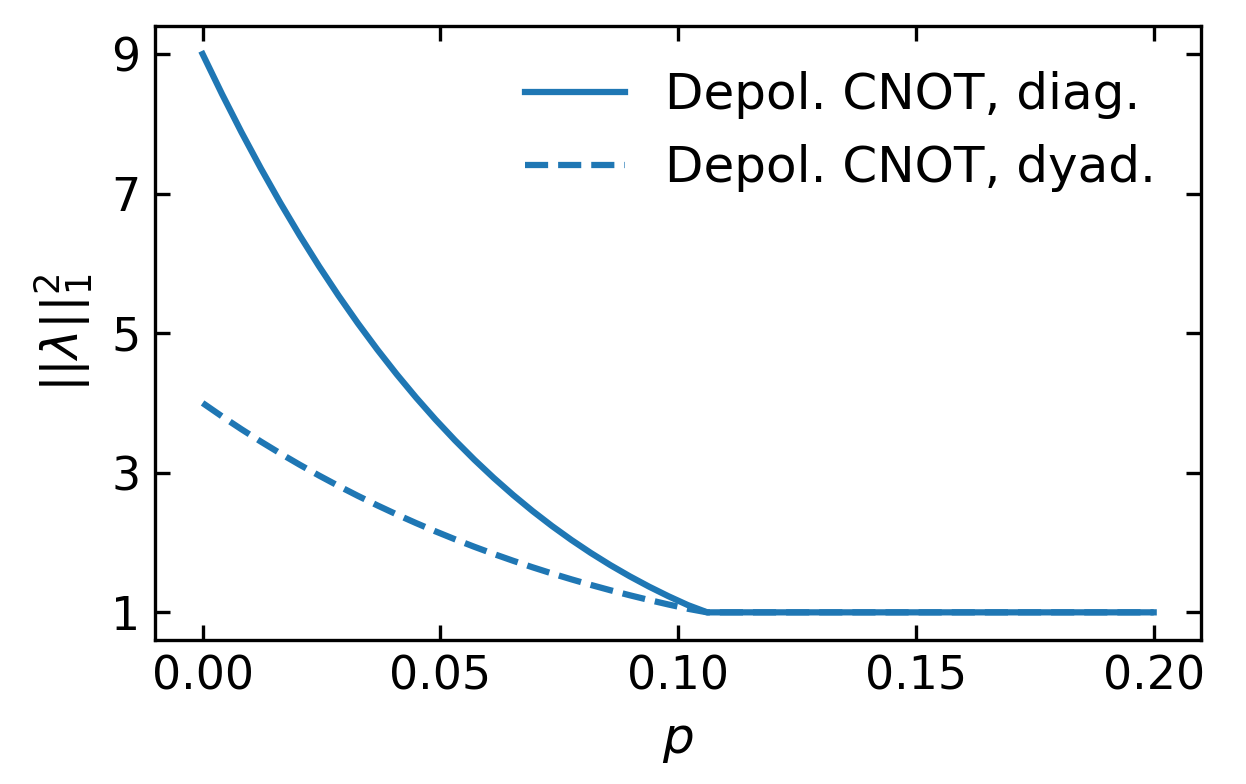}
    \caption{\textbf{Continuous product state frames.} $||\lambda^{(CNOT, \mathrm{depol.})}||_1^{\,2}$ for optimal analytic decompositions with respect to the diagonal (solid line) and dyadic (dashed line) continuous product state frames, plotted as a function of the depolarizing strength $p$.} \label{fig:prod_cont}
\end{figure}

In Fig.~\ref{fig:prod_cont}, we plot this value of $||\lambda^{(CNOT,\rm{depol.})}||_1$ achieved by the continuous dyadic frame and compare it to the one previously derived for the continuous diagonal frame. We observe a behaviour that is completely analogous to the one observed for stabilizer frames: the dyadic frame (dashed line) gives a significant improvement in the simulation cost at low depolarizing strength over the diagonal frame, but the threshold value remains fixed at the same point.

\section{Results in the Heisenberg picture} \label{sec:results-heisenberg}

We now consider algorithms constructed with choices of frames suitable for the Heisenberg picture simulation. Although the general structure of these algorithms is completely analogous to the Schr\"odinger case, our results will show that the particular features of the Heisenberg-adapted frames generally allow them to achieve better performance at comparable noise rates. This appealing feature is due to the fact that these frames are able to overcome the limitations deriving from resource distillation theorems discussed in Appendix~\ref{app:nogo}. This is in particularly true, as we discuss in more detail below, for the Pauli frame methods, which have enjoyed significant attention in the recent literature~\cite{fontana_classical_2025, angrisani_classically_2024, angrisani_simulating_2025, shao_simulating_2024, schuster2024polynomialtimeclassicalalgorithmnoisy, gonzalez-garcia_pauli_2025}. We therefore begin by discussing the Pauli method from the frame theory point of view, giving a new perspective on its good simulation power. We then show that these Pauli-based algorithms are actually still not optimal. We are indeed able to introduce an extension of the Pauli frame which exhibits an improved simulation threshold and is actually optimal within a specific family of frames we discuss below. 

\subsection{Pauli frame} \label{sec:pauli-frame}
If we choose the Pauli operator basis as our frame, we construct a simulation algorithm closely related to the well-known Pauli back-propagation methods. More precisely, we consider here the frame $\mathcal{F}=\{F_x\}_{x\in \cX}$, where the frame elements $F_x=P_x$ are Pauli operators, that is tensor products of the local Pauli matrices $\{\id, X, Y, Z\}$. Each element $F_x$ can therefore be labelled by a string $x$ of $2n$ bits, which provides an efficient parametrisation of the frame. It also follows that the frame size is $|\cX|=4^n$.

First, we notice that $\tr\left[ \rho_0 P_{x}\right]$ can be efficiently computed for $\rho_0=\ket{0^n}\!\bra{0^n}$ and in general for initial states that are stabilizer states or product states. Furthermore equation~\eqref{eq:bounded-expectation-heis} is always satisfied with $ |\tr\left[ \rho_0 P_{x}\right]|\leq 1$ for all $x$. This makes the frame suitable for simulation in the Heisenberg picture. We therefore move on to study equations~\eqref{eq:frame-expansion-state-heis}-\eqref{eq:frame-expansion-gate-heis}, which take the form
\begin{align}
    O&=\sum_{x\in \cX^{(0)}} \lambda_x^{(0)} \, P_x \,, \label{eq:frame-expansion-obs-Pauli}\\
    \mathcal{C}^*_j(P_x)&= \sum_{y\in \cX^{(j)}_x} \lambda_{x,y}^{(j)} \, P_y \,.\label{eq:frame-expansion-gate-Pauli}
\end{align}
The decomposition~\eqref{eq:frame-expansion-obs-Pauli} gives $\|\la_0\|_1=1$ whenever $O$ is a Pauli observable. This corresponds to a trivial decomposition where $\cX^{(0)}$ contains a single element.

Regarding the decomposition of gates~\eqref{eq:frame-expansion-gate-Pauli}, note that the Pauli operators form an orthogonal basis, therefore the decomposition can be uniquely identified as 
\begin{equation}
    \lambda_{x,y}^{(j)} = \frac{1}{2^n}\tr[P_y \, \mathcal{C}^*_j(P_x)] \label{eq:pauli-transfer-matrix}
\end{equation}
and is thus related to the so-called \emph{Pauli transfer matrix} of the channel.
In particular, if $\mathcal{C}^*_j$ is a local channel with support on a constant number of qubits, then the expression above is non-zero only for a constant number of indices $y$, that is the ones such that $P_y$ differs from $P_x$ only on the tensor factors where the gate acts. For these relevant indices $y$, the coefficient $\lambda_{x,y}^{(j)}$ can be evaluated efficiently, as this reduces to a computation in terms of local operators.

Furthermore, if a channel $\mathcal{C}^*_j$ can be written as a convex combination of $\mathrm{poly}(n)$ Clifford channels then~\eqref{eq:frame-expansion-gate-Pauli} takes a form where $\cX^{(j)}_x$ contains polynomially many elements and $\|\lambda^{(j)}_{x}\|=1$. More specifically, consider 
\begin{equation}
\mathcal{C}^*_j(\cdot)=\sum_k p_k C_k^\dagger (\cdot) C_k,
\end{equation}
where $C_k$ are Clifford unitaries. Then, because Cliffords map Pauli operators into Pauli operators, equation~\eqref{eq:frame-expansion-gate-Pauli} is satisfied by a choice where the coefficients $\lambda_{x,y}^{(j)}$ coincide with the positive normalised distribution $p_k$. It follows that this class of channels can be simulated with an efficient algorithm cost. For example, this applies to any local Clifford gate affected by Pauli noise.

On the other hand, channels $\mathcal{C}^*_j$, which do not correspond to a Clifford operation, may in general have $\|\lambda^{(j)}_{x}\|>1$ and contribute to an inefficient simulation cost. Consider for instance a noiseless $T$-gate. We have
\begin{equation}
    T^\dag \, X \, T = \frac{1}{\sqrt{2}}(X+Y),
\end{equation}
which means that, for any Pauli operator $P_x$ which contains an $X$ factor on the qubit where the $T$-gate is acting, $\mathcal{C}^*_j(P_x)=(P_x + P'_x)/\sqrt{2}$. Here $P_x'$ is the same Pauli string as $P_x$, where the $X$ factor was replaced by a $Y$. This corresponds to a decomposition~\eqref{eq:frame-expansion-gate-Pauli} where $\lambda_{x,y}^{(j)}$ takes two different non-zero values and $\|\lambda^{(j)}_{x}\|=\sqrt{2}>1$.

\begin{figure}
    \centering
    \includegraphics[width=\linewidth]{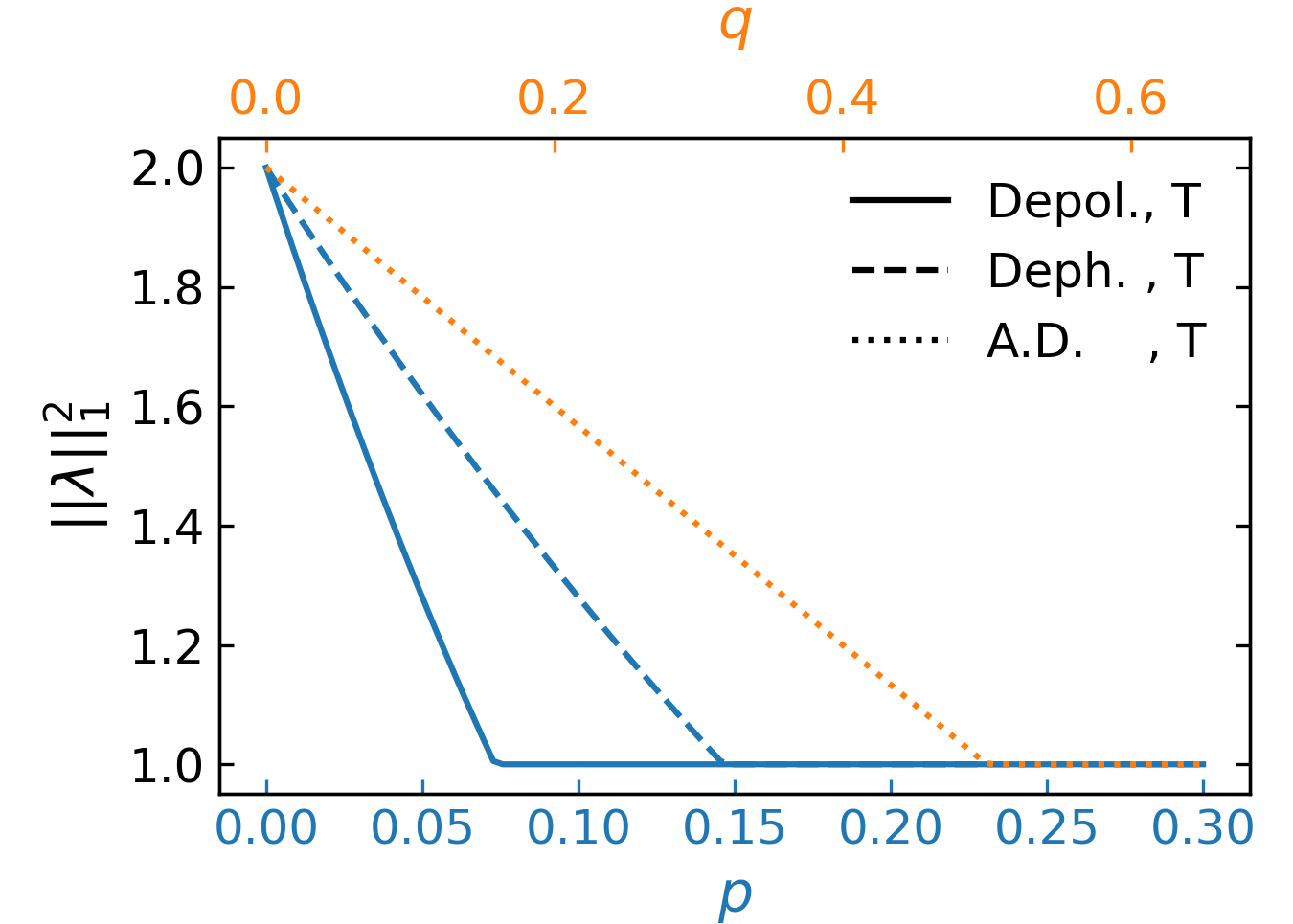}
    \caption{\textbf{Pauli frame.} $||\lambda^{(j)}||_1^{\,2}$ for decompositions in the Pauli frame of the noisy $T$-gate at different depolarizing, dephasing and amplitude damping (A.D.) strengths. Note that the depolarizing and dephasing strengths $p$ and the amplitude damping strength $q$ are plotted in separate axes as they take values in different ranges.  }
    \label{fig:pauli}
\end{figure}

In the case of noisy non-Clifford gates, the specific value of $\|\lambda^{(j)}_{x}\|$ will depend on the interplay between the unitary gate and the strength of the noise. In Figure~\ref{fig:pauli} (solid line), we plot the value of $\|\lambda^{T, {\rm  depol.}}\|_1$ for the channel $\cC_{T,{\rm  depol.}}(\cdot)=\mathcal{N}^{\rm  depol.}_p(T (\cdot) T^\dag)$, that is a $T$-gate followed by depolarising noise of strength $p$, as defined in~\eqref{eq:depol}. Note that here, as the Pauli frame is also a basis, there is no need to optimize over possible channel decompositions: the plotted results are derived directly by computing~\eqref{eq:pauli-transfer-matrix} for the given channel. We observe also in this case an inverse threshold behaviour. In the Pauli frame, the Clifford+$T$ gate-set becomes classically simulable above a critical noise strength $p_{\rm cl}\approx 0.07$, considerably lower than the corresponding threshold in the stabiliser-based frames discussed in Section~\ref{sec:stabiliser-frames}.

It may seem surprising, given the discussion in Appendix~\ref{app:nogo}, that the Pauli frame algorithm is able to efficiently simulate a noisy $T$-gate at lower noise rates than stabilizer frames (\textit{i.e.} below the magic distillation threshold $p_{\rm dist}\approx 0.11$). However, there is no contradiction because, as previously noticed by Rall \textit{et al.}~\cite{rall_simulation_2019}, the Pauli frame method is not able to efficiently simulate adaptive Clifford operations and thus the theorems of Appendix~\ref{app:nogo} do not apply. Indeed, any adaptive gate has $\|\lambda\|_1>1$ in this frame. In conclusion, it is precisely this limitation of the Pauli method (not being able to simulate adaptive circuits), that at the same time allows it to circumvent the no-go theorem and reach a lower simulability threshold.

We conclude by noting that the same observations can be repeated for other noise models. As examples, in Figure \ref{fig:pauli} we also plot the case of the dephasing noise channel~\eqref{eq:deph} (dashed line) and the amplitude damping channel (dotted line), which is defined as
\begin{align}
    \mathcal{N}^{\rm  A.D.}_q(\rho):= K_1 \rho K_1^\dag + K_2 \rho K_2^\dag   ,\label{eq:AD} 
\end{align}
with Kraus operators 
\begin{equation}
K_1 =\begin{pmatrix}
      1 & 0\\
      0 & \sqrt{1-q}
     \end{pmatrix}, 
K_2 = \begin{pmatrix}
      0 & \sqrt{q}\\
      0 & 0
     \end{pmatrix}.
\end{equation}
Notice in particular that the latter is a non-unital channel, representing a typically more challenging form of noise to simulate than unital Pauli channels (even in average-case settings~\cite{mele_noise-induced_2024, angrisani_simulating_2025}). It is not immediately obvious that Clifford gates perturbed by amplitude damping noise can always be simulated efficiently, however this can also be separately verified with the same analysis for this frame.

\subsection{Extended Pauli frame} \label{sec:extended-pauli-frame}
So far, for what concerns frames applied in the Heisenberg picture, we have only considered a frame (the Pauli frame) which is also a basis. This in particular means that the decomposition~\eqref{eq:pauli-transfer-matrix} is unique and the corresponding norm $\|\lambda\|_1$ directly determines the runtime of the resulting algorithm. However, we have seen in the Schr\"odinger picture section that it is often advantageous to exploit overcomplete frames, where the existence of many possible decompositions allows us to choose the one with the optimal one-norm value. We will now therefore discuss how to extend the Pauli basis to an overcomplete frame where one can find decompositions with a better one-norm value than in the Pauli case.

We find that a promising choice is given by the frame $\mathcal{F}=\{F_x\}_{x\in\cX}$ where
\begin{align}
    F_x=\bigotimes_{k=1}^n f_x^{(k)}\,.\label{eq:generalized-pauli-frame}
\end{align}
Here the tensor product runs over all the qubits in the system and $f_x^{(k)}$ are local operators on the $k$-th qubit, chosen from the following set
\begin{align}
    f_x^{(k)}&\in\left\{\id,\,X,\,Y,\,Z,\, \frac{a}{\sqrt{2}}(X+Y),\, \frac{a}{\sqrt{2}}(X-Y)\right\}\,, \label{eq:generalized-pauli-set}
\end{align}
for a fixed choice of the hyperparameter $a\in[0,1]$. To fully identify a frame element $F_x$, we need to specify each operator $f_x^{(k)}$ among this constant number of possible choices. This means that the frame can be efficiently parametrised by a string $x$ of $O(n)$ bits.
Notice that this is a direct generalization of the Pauli frame, which corresponds to neglecting the last two elements of the set~\eqref{eq:generalized-pauli-set}, \ie setting $a=0$. 

First of all, we observe that, just as in the Pauli case, we still have that $|\tr\left[ \rho_0 F_{x}\right]|\leq ||F_x||\leq 1$ for all $x$ and that $\tr\left[ \rho_0 F_{x}\right]$ can be efficiently computed for $\rho_0=\ket{0^n}\!\bra{0^n}$ or any product initial state. Furthermore, if we focus on Pauli observables $O$, there exists always a decomposition~\eqref{eq:frame-expansion-state-heis} with $\|\la_0\|_1=1$ (which just coincides with the decomposition on the Pauli basis).

For what concerns the decomposition of gates, we instead encounter a different scenario than in the Pauli case. Indeed, it is no longer true that any convex combination of Clifford unitaries automatically has $\|\lambda^{(j)}\|_1\leq 1$, as the frame elements $a(X\pm Y)/\sqrt{2}$ are not mapped to other frame elements. We thus need to carefully examine also the case of Clifford gates in our following analysis. 

Consider therefore a generic channel $\cC_j$ acting on a constant number $q$ of qubits (in the examples that follow this will be at most $q=2$). If, for instance, it acts non-trivially only on the first $q$ qubits, we can write it as $\cC_j=\cC'_j\otimes \mathcal{I}^{n-q}$ (for gates acting on other subsets of qubits the discussion follows analogously). We now need to find efficient choices of the coefficients $\lambda^{(j)}_{x,y}$ such that 
\begin{align}
    \mathcal{C}^*_j(F_x)&= \sum_{y\in \cX} \lambda_{x,y}^{(j)} \, F_y  \label{eq:frame-expansion-gate-gen-Pauli}
\end{align}
holds for each $x$. By comparing this with the frame definition~\eqref{eq:generalized-pauli-frame}, it is clear that it makes sense to consider decompositions with support on those frame elements $F_y$ that differ from $F_x$ only in the factors $f_x^{(k)}$ where the gate is acting (\ie for $k\leq q$). With this choice, equation~\eqref{eq:frame-expansion-gate-gen-Pauli} reduces to 
\begin{align}
    {\mathcal{C}'_j}^*\left(\bigotimes_{k=1}^q f_x^{(k)}\right)&= \sum_{y\in \cX^{(j)}_x} \lambda_{x,y}^{(j)} \, \bigotimes_{k=1}^q f_y^{(k)}\,, \label{eq:frame-expansion-gate-gen-Pauli-reduced}
\end{align}
where we have defined $\cX^{(j)}_x$ as precisely those frame elements which differ from $F_x$ only in the first $q$ factors. As the size of the set~\eqref{eq:generalized-pauli-set} and $q$ are both constant, also the size of $\cX^{(j)}_x$ is constant. Furthermore, equation~\eqref{eq:frame-expansion-gate-gen-Pauli-reduced} now relates operators on a system of constant size $q$. It is therefore possible to efficiently characterize the set of $\lambda_{x,y}^{(j)}$ which satisfy it, independently of the system size $n$, and to find the choice which minimizes $\|\lambda^{(j)}\|_1$.

In particular, we proceed again as we did in Section~\ref{sec:optimization-problem} for stabiliser frames. Let $y \in \mathbb{C}^{4^q}$ be the vector representation of the left-hand side of equation~\eqref{eq:frame-expansion-gate-gen-Pauli-reduced} in the Pauli basis of $q$-qubit operators, and consider the matrix $D\in\mathbb{C}^{4^q\times|\cX^{(q)}|}$, whose columns are the vector representations in the same Pauli basis of all $q$-qubit frame elements $\bigotimes_{k=1}^q f_x^{(k)}$. Here, $|\cX^{(q)}|$ is the dimension of this $q$-qubit frame, which is constant. Then, for all the values of $y$, corresponding to left-hand side of~\eqref{eq:frame-expansion-gate-gen-Pauli-reduced} for each choice of $\bigotimes_{k=1}^q f_x^{(k)}$, we need to solve the optimization problem
\begin{equation} \label{eq:opt-prob-ext-pauli}
\begin{aligned}
\min_{\lambda \in \mathbb{C}^{|\cX^{(q)}|}} \quad & ||\lambda||_1 \\
\textrm{s.t.} \quad & D \lambda = y.
\end{aligned}
\end{equation}
As discussed previously this can be mapped to the form~\eqref{P2} and solved efficiently in the dimension of the problem (which is $4^q\times|\cX^{(q)}|=4^q\times 6^q$).

\begin{figure}
  \includegraphics[width = \linewidth]{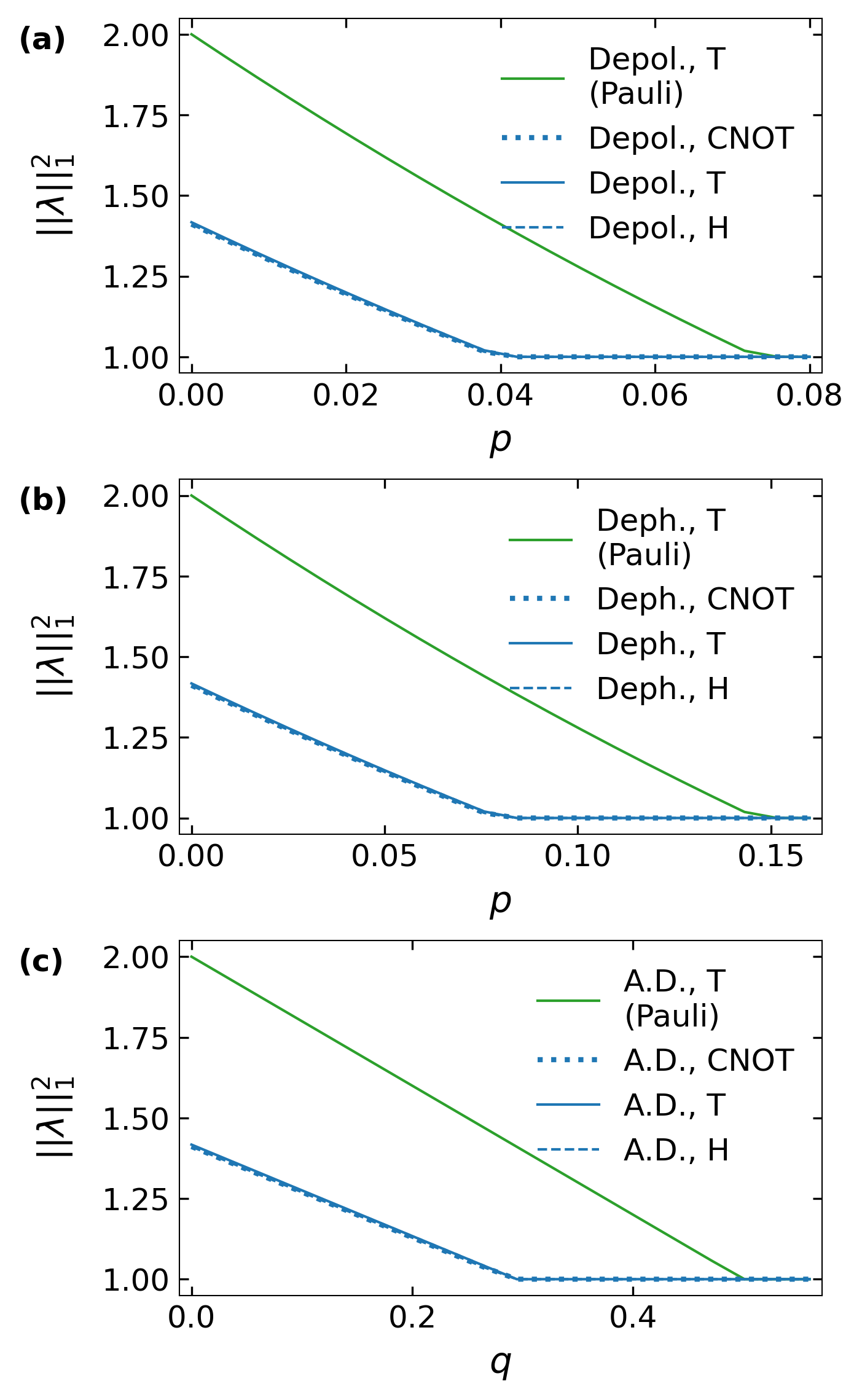}
  \caption{\textbf{Extended Pauli frame.} $||\lambda^{(j)}||_1^{\,2}$ for optimal decompositions in the extended Pauli frame at different noise strengths (blue lines). We assume a universal gate-set composed of noisy Hadamard ($H$), controlled-NOT ($CNOT$) and magic ($T$) gates. The noise is given by the depolarizing (panel (a)), dephasing (panel (b)) and amplitude damping (panel (c)) channels. The extended Pauli frame is defined with the choice $a=0.84$. For comparison we also plot the value of $||\lambda^{(T)}||_1^{\,2}$ for the standard Pauli frame (green line).}
  \label{fig:ext_pauli_depol}
\end{figure}

In Figure~\ref{fig:ext_pauli_depol}, we plot the results of this optimization for noisy gate-sets composed of the Hadamard ($H$) and controlled-NOT ($CNOT$) Clifford gates and the standard magic gate ($T$), all followed by either depolarizing, dephasing or amplitude damping noise (respectively the noise channels~\eqref{eq:depol}, \eqref{eq:deph} or~\eqref{eq:AD}). 
First of all, note that the outcome of this analysis depends on the specific choice of the hyperparameter $a$ in the definition of the frame. Choosing a lower value for $a$ makes the frame closer to the Pauli frame, where the Clifford gates are always efficiently simulable and the $T$-gate is not. Vice-versa, a higher value of $a$ makes the simulation of the $T$-gate easier (as objects such as $T^\dagger XT$ are now approximately contained in the frame), at the cost of making the Clifford gate $CNOT$ less efficient. There exists an optimal value of $a$ which balances this trade-off and achieves the optimal global threshold value for the complete gate-set. We empirically find this value to be $a\approx 0.84$, which corresponds to the results show in the figure. We observe that, with this choice of $a$, the extended Pauli frame gives a significantly improved simulation cost and a lower inverse threshold compared to the standard Pauli basis, at least for these gate models.

For the specific choice of noise channels $\mathcal{N}^{\rm  depol.}_p$, $\mathcal{N}^{\rm  deph.}_p$ and $\mathcal{N}^{\rm  A.D.}_q$, we are able moreover to find analytical solutions to the optimization problem~\eqref{eq:opt-prob-ext-pauli}. We discuss these more in detail in Appendix~\ref{app:analytics-ext}, further proving that the value achieved for these noise models by the extended Pauli frame is optimal among all frame choices of the form~\eqref{eq:generalized-pauli-frame}, where $f_x^{(k)}$ are operators with $\|f_x^{(k)}\|\leq 1$.

\section{Outlook}\label{sec:outlook}

\subsection{A frame-based perspective on the boundary between the quantum and classical worlds}

The frame-based perspective developed in this work opens up a 
wealth
of directions for future research, both on the conceptual and on the algorithmic side. A central property of this framework is its flexibility: once a simulation problem is cast in terms of frames and quasi-probabilities, the choice of frame directly determines the simulation cost through the value $\|\lambda\|_1$. Operations for which this value is one are ``free'' and can be simulated efficiently classically. Larger values instead contribute to an exponential simulation cost and identify regimes where advantage may be possible.

For some of the frames discussed in this paper, $\|\lambda\|_1$ is related to known quantities (\textit{e.g.} the \emph{robustness of magic} for the diagonal stabilizer frame and the \emph{robustness of entanglement} for the product state frame), while for other frames this is a genuinely novel simulability quanitifier.
This approach allows a larger freedom to tackle a broader class of problems than the standard tools for ``easing'' the sign problem of Monte Carlo methods \cite{Easing,Signs1, SignsTerhal}.

\subsection{Novel directions for frames to be explored}

In general, it is very natural to explore frames associated with other notions of quantum resources beyond those explicitly considered here. In fact, our work provides a general theory of frame simulation, giving clear prescriptions on the properties that new frame candidates must satisfy. This now makes frame discovery a well-defined and approachable task. While some frames introduced here, such as the extended Pauli frame, are provably optimal within their family, there is still ample scope for developing new frame families. Indeed, as we have shown for Heisenberg picture simulation, one can substantially improve on known results by more elaborate frame choices, highlighting how subtle it is to precisely delineate the boundary between the classical and quantum settings.

A particularly promising direction 
is to move beyond product-state frames toward frames built from weakly entangled states, such as matrix product states with bounded bond dimension. This would provide a controlled interpolation between mean-field–type simulations and those based on area-law states, making a connection with the rich literature on MPS-based classical simulation algorithms~\cite{gao_efficient_2018, vovk2024, chen_optimized_2024, sander_large-scale_2025, cichy2025classicalsimulationnoisyquantum}.
In these works, decomposition of the state are sought which minimize various measures of entanglement. Our approach would provide a different and potentially fruitful perspective where instead the focus is on the suitable one-norm quantities, according to the Monte Carlo analysis.

Similarly, frames composed of Gaussian states could be used to study the role of non-Gaussianity in noisy circuits, while frames adapted to coherence -- where diagonal gates in a fixed basis are free and basis-changing gates such as Hadamard are resourceful -- could offer a complementary view on coherence-based resource theories. Finally, the well-known discrete Wigner representation \cite{StabilizerPolytope,Wigner} 
also fits naturally into our formalism, providing an example of an operator basis suited for Schr\"odinger rather than Heisenberg picture simulation.

Looking ahead, another possible extension is the use of adaptive frames, where the frame itself is allowed to change dynamically during the simulation, depending on the gates applied or on intermediate sampling outcomes. This is analogous to the ``dynamical easing of the sign problem'' of standard Monte Carlo approaches.
Such adaptivity could potentially reduce accumulation of sampling cost and lead to substantially improved algorithms, at the price of a more complex classical control structure. Developing systematic principles for choosing or updating frames on the fly is an open challenge.
In summary, we emphasize that the frame viewpoint provides a common language for a wide class of existing and yet-to-be-developed simulation strategies.

\subsection{The worst-case nature of estimates}

Another important aspect of our results is that they apply to concrete circuits rather than only to ensembles. The bounds derived from the frame formalism are worst-case guarantees: for any fixed circuit composed of gates from a given noisy gate-set, one can rigorously bound the simulation cost in terms of the associated one-norm. Furthermore, even when the one-norm is not known exactly, the empirical variance of the Monte Carlo estimator provides a rigorous, a posteriori diagnostic of performance. This means that for any individual circuit one can efficiently certify whether the simulation succeeded and the reported estimates are reliable. This stands in contrast to some of the recent literature on noisy circuits focusing on \emph{average-case simulability}, where the quantities needed to certify the success of the simulation are not always efficiently accessible~\cite{mele_noise-induced_2024,angrisani_simulating_2025}. We anticipate this to be a powerful feature of our methods, enabling their application in a wide range of practical scenarios, where one can expect the one-norm $\|\lambda\|_1$ to represent a simple but rather pessimistic upper-bound on the actual performance of the algorithm.

For example, our extended Pauli frame results give an improved value for the inverse threshold of circuits compiled from the unitary Clifford+$T$ gate-set where each gate is followed by depolarizing noise, with no further assumption on the  circuit architecture or on the supply of fresh (\textit{i.e.} noiseless) ancilla qubits. This is a class of circuits which is known to contain instances that encode fault tolerant universal quantum computation when the noise rate is low enough~\cite{AB08, MeaseurementFree}. Our method allows to explore it in detail, in contrast to many other common simulation techniques which rely either on average case statements over random circuit ensembles, or on the assumption that all qubits are affected  by the noise after each computation layer.

\subsection{Further practical and technological applications}

Finally, by treating both the Schrödinger and the Heisenberg pictures, our work also contributes to clarify the relationship between different notions of simulation. While our framework is specifically tailored to the estimation of expectation values, there are specific scenarios where it can also be used for weak simulation (\ie the sampling of measurement outcomes). This is particularly the case in the Schrödinger picture simulation using frames of the form $F_x=\ket{\psi_x}\!\bra{\psi_x}$, for a suitable set of states $\psi_x$. Here, an efficient simulation regime necessarily also corresponds to a convex representation of the evolved state in the frame, which can be used to perform weak simulation. In this context, it may also be fruitful to consider instances of ``interaction picture'' representations between the Schrödinger and the Heisenberg ones.

The framework presented here may find natural applications in benchmarking and verification of near-term quantum devices \cite{BenchmarkingReview,SystematicBenchmarking,EUBenchmarking}.
It can indeed provide classically computable reference values for the outputs of circuits with sufficiently strong noise or even of the average outputs over ensembles of noiseless circuits, which can then be used to test device performance~\cite{baccari2025averagecomputationbenchmarkinglocalexpectation}. In this context it is interesting to note that the formalism extends naturally to study higher moments and collective measurements. In particular, by considering tensor products of the analysed frames, one can straightforwardly compute multi-copy quantities that can be written as expectation values with respect to $\rho^{\otimes q}$ for small constant $q$ -- \textit{e.g.} purities. This paves the way to explore more advanced applications, such as monitored circuits, computations with intermediate measurements and the injection of fresh qubits. These play an increasingly important role in both experimental platforms and theoretical studies of quantum dynamics and their study will be a direction for future work.

Moving away from the circuit model, a further interesting question is to apply these methods to the dynamics of many-body systems. Here, the natural scenario would be to consider local Liouvillian dynamics, where the few-body nature of each individual Liouvillian term potentially allows for an optimization over frame representations similar to the ones considered in this work.

\section{Conclusion}
In this work, we have studied a unified framework for the classical simulation of noisy quantum circuits grounded in frame theory, which consolidates diverse simulation strategies within a single mathematical formalism. Simulation cost being rigorously linked to the one-norm of quasi-probability representations, this approach enables 
direct performance comparisons across different Schrödinger- and Heisenberg-picture methods. Exploring different frame choices led us to new and improved simulation schemes, like a generalized Pauli frame and to refined bounds on simulation efficiency that hold for individual circuits and fixed gate-sets. These results establish a rigorous, worst-case approach for deriving inverse thresholds and highlight frame theory as a powerful tool for both understanding and advancing the limits of classical simulability. On a higher level, this work helps to delineate the boundary between the noisy quantum circuits that can still be classically simulated and those for which this is no longer possible and one can hope quantum advantages can set in. This provides a tool to assess whether \emph{noisy intermediate scale quantum} (NISQ) computers have the potential to feature quantum utility, or whether the step to a full \emph{fault-tolerant application-scale quantum} (FASQ) \cite{MindTheGaps} computer will eventually be necessary to arrive at robust quantum utility.

\section{Acknowledgements}
JD acknowledges the support of MUNIQC-Atoms and the Quantum Flagship PASQuanS2.
JC acknowledges the support of Berlin Quantum.
JE acknowedges support of the BMFTR (Hybrid++, QuSol), the Munich Quantum Valley, Berlin Quantum, QuantERA, the Quantum Flagship (Millenion, PASQuanS2), The Clusters of Excellence (MATH+, ML4Q), and the European Research Council (project DebuQC, grant agreement No.\ 101098279). 
TG acknowledges the support by the European Research Council (project DebuQC, grant agreement No.\ 101098279).

\appendix

\section{Dual formulation of optimal representations} \label{app:duals}
\label{sec:appendixA}
In this Appendix, we provide an alternative characterization of the problem of finding optimal frame representations of a given gate, by introducing the \emph{dual} optimization problem. This formulation will then be used extensively in Appendix~\ref{app:analytics} to derive analytical expressions for our main results. For completeness, we provide here a self-contained mathematical presentation of the necessary notions, although these are well-established in the literature~\cite{Boyd,seddon_quantifying_2021}.

More specifically, the following Theorem~\ref{thm1} provides a dual characterization of the optimal coefficient one-norm of an operator in a given frame. In all applications below, the inner product space $V$ is either the complex vector space of linear operators $\mathcal{L}(\mathcal{H})$ on some finite-dimensional Hilbert space $\mathcal{H}$, or the real vector space of Hermitian operators on $\mathcal{H}$, both equipped with the Hilbert-Schmidt inner product.

\begin{theorem}[Strong duality]
\label{thm1}
Let $(V,\Braket{\cdot|\cdot})$ be a finite-dimensional inner product
space over $\mathbb{R}$ or $\mathbb{C}$, and let
$\mathcal{F}=\{F_x\}_{x\in \mathcal{X}}\subset V$ be a (bounded) frame.
Let
\begin{equation} \label{eq:def-frame-norm}
\left\lVert A \right\rVert_{\mathcal{F}}
:=
\inf\left\{
\left\lVert \lambda \right\rVert_{1}:
A =\sum_{x\in\mathcal{X}} \lambda_x F_x
\right\}.
\end{equation}
Then, for all $A \in V$,
\begin{align} \label{eq:dual-frame-norm}
\left\lVert A \right\rVert_{\mathcal{F}}
=
 & \max\left\{
  \Braket{W|A}:  
 W\in V, \right.  \\
& \qquad\qquad \left. 
\left\lvert \Braket{W|F_x}\right\rvert\leq 1, \,  \forall x \in \mathcal{X} \right\}. \nonumber
\end{align}
\end{theorem}
In particular, this statement of strong duality lets us certify the optimality of a given decomposition into frame elements. From the definition~\eqref{eq:def-frame-norm}, we have that any valid decomposition with coefficients one-norm $\|\lambda\|_1$ provides an upper bound to the optimal value $\left\lVert A \right\rVert_{\mathcal{F}}$. Vice versa, from~\eqref{eq:dual-frame-norm} we find that any operator $W$ that satisfies $|\Braket{W|F_x}|\leq 1$ for all frame elements (in what follows we call such operators \emph{dual-feasible} certificates), gives a lower bound $\Braket{W|A}$ for the same quantity $\left\lVert A \right\rVert_{\mathcal{F}}$. If one finds a dual-feasible certificate $W$, giving a matching bound to the coefficient one-norm $\|\lambda\|_1$ of a given decomposition, then $W$ certifies the optimality of said decomposition.

As we will see in Appendix~\ref{sec:appendixB}, finding an optimal $W$ for the types of frames we study in this paper can be done either by close inspection or simple computation. Finally, before delving into the proof of Theorem~\ref{thm1}, let us remark that, besides trying to find optimal decompositions with respect to a specific frame, this duality also allows us to explore the limits of certain \emph{classes} of frames. Indeed, in Appendix~\ref{sec:appendixB}, we show that no properly normalized product frame can do better than the extended Pauli frame we introduced in \eqref{eq:generalized-pauli-set}, assuming local noise that can have depolarizing, dephasing, or even amplitude damping components.

\begin{proof}

We show both inequalities separately.

\medskip
\noindent
\textbf{``$\geq$''.}
Given a decomposition
\begin{equation}
A=\sum_x \lambda_xF_x \,,
\end{equation}
the triangle inequality establishes
\begin{equation}
\Braket{W|A}
\leq
\sum_x \left\lvert \lambda_x\right\rvert
\left\lvert \Braket{W|F_x}\right\rvert.
\end{equation}
If $W$ satisfies $\left\lvert \Braket{W|F_x}\right\rvert\leq 1$ for
every $x$, then
\begin{equation}
\Braket{W|A}
\leq
\sum_x\left\lvert \lambda_x\right\rvert.
\end{equation}
This holds for arbitrary decompositions, and therefore
\begin{equation}
\Braket{W|A}
\leq
\left\lVert A\right\rVert_{\mathcal{F}}.
\end{equation}
Taking the supremum over all dual-feasible $W$ gives the desired
inequality. As we shall see shortly, the supremum is in fact attained by a specific choice of $W$.

\medskip
\textbf{``$\leq$''.}
The case $A=0$ is immediate, so assume $A \neq 0$ and set
\begin{equation}
A'=\frac{A}{\left\lVert A \right\rVert_{\mathcal{F}}}.
\end{equation}
Then, define the closure
\begin{equation}
\mathcal{C}
=
\begin{cases}
\overline{\operatorname{conv}}\{\pm F_x: x\in \mathcal{X}\},
& V\text{ real},\\[2mm]
\overline{\operatorname{conv}}\{e^{i\varphi}F_x:
       x\in\mathcal{X},\ \varphi\in[0,2\pi)\},
& V\text{ complex}.
\end{cases}
\end{equation}
The set $\mathcal{C}$ contains all convex combinations of frame elements, up to phases, and is the unit ball of the norm
$\left\lVert \cdot\right\rVert_{\mathcal{F}}$, so $A'\in\partial\mathcal{C}$ lies on the boundary.

In Lemma \ref{lem:supp_hyperplane}, we prove a variant of the supporting hyperplane theorem \cite{Boyd}, which implies the existence of a $W\in V$ such that
\begin{equation}
\Braket{W|A'}=1,
\qquad
\left\lvert \Braket{W|X}\right\rvert\leq 1
\quad\text{for all }X\in\mathcal{C}.
\end{equation}
In particular, said $W$ is dual-feasible, that is, the absolute value of its inner product with all frame elements is bounded by $1$. Moreover,
\begin{equation}
\Braket{W|A}
=
\left\lVert A \right\rVert_{\mathcal{F}}\Braket{W|A'}
=
\left\lVert A \right\rVert_{\mathcal{F}},
\end{equation}
which proves the reverse inequality.
\end{proof}

\begin{lemma}[Supporting hyperplane]
\label{lem:supp_hyperplane}
Let $V$ be a finite-dimensional inner product space over
$\mathbb{R}$ or $\mathbb{C}$,
$\{F_x\}_{x\in\mathcal{X}}$ a (bounded) frame. Define
\begin{equation}
\mathcal{C}
=
\begin{cases}
\overline{\operatorname{conv}}\{\pm F_x: x\in \mathcal{X}\},
& V\text{ real},\\[2mm]
\overline{\operatorname{conv}}\{e^{i\varphi}F_x:
       x\in\mathcal{X},\ \varphi\in[0,2\pi)\},
& V\text{ complex}.
\end{cases}
\end{equation}
If $Z_0\in\partial\mathcal{C}$, then there exists a $W\in V$
such that
\begin{equation}
\Braket{W|Z_0}=1,
\qquad
\left\lvert \Braket{W|X}\right\rvert\leq 1
\quad\text{for all }X\in\mathcal{C}.
\end{equation}
\end{lemma}

We note that $\operatorname{Re}(\Braket{W|X}) \leq 1$ defines a supporting hyperplane of the unit ball $\mathcal{C}$, touching $\mathcal{C}$ at $Z_0$, with the entire set on one side.
\begin{proof}
First, we take a $Z\notin\mathcal{C}$ and prove the existence of a suitable $W$ in the case of an orthogonal projection onto $\mathcal{C}$. Later, we will consider a sequence that converges to a general boundary element $Z_0$ from outside $\mathcal{C}$.

Since $\mathcal{C}$ is by definition compact, there exists a
$Z^*\in\partial\mathcal{C}$ that is closest to $Z$ with
respect to the norm induced by the inner product on $V$.
Define the shortest vector connecting $Z$ to $\mathcal{C}$ as
\begin{equation}
W'=Z-Z^*,
\end{equation}
and set
\begin{equation}
Z_\theta^*=\theta X+(1-\theta)Z^*\in\mathcal{C},
\end{equation}
for arbitrary $X\in\mathcal{C}$ and $\theta\in[0,1]$.
By our choice of $Z^*$,
\begin{equation}
\left\lVert Z-Z^*\right\rVert
\leq
\left\lVert Z-Z_\theta^*\right\rVert,
\end{equation}
or equivalently
\begin{equation}
\left\lVert W'\right\rVert
\leq
\left\lVert W'+\theta(Z^*-X)\right\rVert.
\end{equation}
Squaring both sides and looking at the limit $\theta \to 0^+$ gives
\begin{equation}
0\leq\operatorname{Re}\Braket{W'|Z^*-X}.
\end{equation}
In fact, since the set $\mathcal{C}$ is phase invariant, we have that the relation above must hold for all $e^{i\phi}X$ (or all $\pm X$ in the real case) and $\braket{W'|Z^*}\in\mathbb{R}$ (as by definition $\Braket{Z|Z^*} \geq 0$). Hence, the stronger inequality
\begin{equation}
\left\lvert \Braket{W'|X}\right\rvert
\leq
\Braket{W'|Z^*}
\qquad
\forall X\in\mathcal{C}
\end{equation}
must also hold.
We may therefore rescale $W'$ by $1/\Braket{W'|Z^*} > 0$, in order to prove the statement for a general orthogonal projection $Z^* \in \partial\mathcal{C}$:
\begin{equation}
\label{eq:supp_hyperplane_for_proj}
\left\lvert \Braket{W'|X}\right\rvert
\leq
\Braket{W'|Z^*} = 1
\qquad
\forall X\in\mathcal{C} 
\end{equation}

\medskip
\noindent
\textbf{Passing to the boundary.}
Now, let $Z_0\in\partial\mathcal{C}$, and construct the convergent sequence
\begin{equation}
Z_k=\left(1+\frac1k\right)Z_0
\notin\mathcal{C},
\qquad
Z_k\longrightarrow Z_0.
\end{equation}
For each $k$, choose $W_k'$ like above,
and let $Z_k^*$ denote an orthogonal projection of $Z_k$ onto $\mathcal{C}$.

Suppose for the moment that the sequence $W_k'$ is bounded. It then has a convergent subsequence
$W_{k_n}'$.
By continuity of the inner product, denoting the limit of this subsequence with $W$,
\begin{align}
\Braket{W|Z_0}
& =
\lim_{n\to\infty}
\Braket{W_{k_n}'|Z_{k_n}} \\
& =
\lim_{n\to\infty}
\Braket{W_{k_n}'|Z_{k_n}^*} 
=
1,
\end{align}
and
\begin{align}
&\left\lvert \Braket{W|X}\right\rvert
-\left\lvert \Braket{W|Z_0}\right\rvert\\
&= 
\lim_{n\to\infty}
 \Bigl(
\left\lvert \Braket{W_{k_n}'|X}\right\rvert
-
\left\lvert \Braket{W_{k_n}'|Z_0}\right\rvert
\Bigr)\\
&\leq 
\lim_{n\to\infty} 
\Bigl(
\left\lvert \Braket{W_{k_n}'|X}\right\rvert
-
\left\lvert \Braket{W_{k_n}'|Z_{k_n}^*}\right\rvert
\\ \nonumber 
& \hspace{30mm}+ \left\lvert \Braket{W_{k_n}'|Z_0-Z_{k_n}^*}\right\rvert
\Bigr) \\ 
& \leq 0,
\end{align}
where we used the inverse triangle inequality.

\medskip
\noindent
\textbf{Boundedness of $W_k'$.}
It remains to justify the boundedness used above.
Due to the equivalence of norms in finite dimensions, there exists $r>0$ such that
\begin{equation}
\{X\in V:\left\lVert X\right\rVert\leq r\}\subseteq\mathcal{C}.
\end{equation}
In particular,
\begin{equation}
r\frac{W_k'}{\left\lVert W_k'\right\rVert}
\in\mathcal{C}.
\end{equation}
Using Inequality~\eqref{eq:supp_hyperplane_for_proj}, we get
\begin{equation}
\left\lvert
\Braket{W_k'|
r\frac{W_k'}{\left\lVert W_k'\right\rVert}
}
\right\rvert
= r \left\lVert W_k'\right\rVert \leq 1,
\end{equation}
which proves boundedness and completes the proof.
\end{proof}

\section{Analytical values of simulation thresholds and optimal decompositions} \label{app:analytics}
\label{sec:appendixB}
In this Appendix, we derive the analytical values of the inverse simulation thresholds for some of the models considered in the main text. While the numerical analysis presented in the main text shows that our approach can be efficiently applied to a large class of gate-sets and noise models, the results presented here give a more precise characterization of some specific and interesting cases.
 
\subsection{Stabilizer frame} \label{app:analytics-stab}
In Appendix~\ref{app:nogo}, we present an argument why the stabilizer frame simulation algorithms cannot be efficient for depolarization strengths lower than $p_{\rm dist}=(6-2\sqrt{2})/28$. Here, we show that these algorithms are optimal, in the sense that they are efficient for any $p\geq p_{\rm dist}$. These two observations combined establish that the inverse threshold for the noisy gate $\mathcal{C}_{T,\rm{depol.}}$ is indeed exactly $p_{\rm cl}=p_{\rm dist}=(6-2\sqrt{2})/28\approx 0.11$, consistently with the numerical results presented in Figure~\ref{fig:stabs_depol}.

To show that stabilizer frame algorithms are efficient for $p\geq (6-2\sqrt{2})/28$, we rely on the fact that above this threshold the noisy T-gate $\mathcal{C}_{T,\rm{depol.}}$ can be written as a convex combination $\sum_k p_k C_k (\cdot) C_k^\dag$ of Clifford unitary channels. This was proven in Ref.~\cite{buhrman_new_2006}, where it was shown that for these noise values the matrix $(1-4p)R_z(\pi/4)$ ---corresponding to the noisy T-gate action on the Bloch vector representation of single qubit states--- is contained in the convex polytope generated by the 24 rotation matrices representing single-qubit Clifford unitaries. An explicit form of this decomposition can be derived, but it is not particularly informative and we omit it here.
As argued in Section~\ref{sec:stabiliser-frames-diag}, convex combinations of Clifford gates are always efficiently simulable in the stabilizer frames. 

As a side note, let us stress that, in general, efficient classical simulability with the stabilizer frames is not equivalent to the gate channels being convex combinations of Clifford unitaries. In principle, there may exist channels that are not themselves convex Clifford combinations, yet map all stabilizer states to convex combinations of stabilizer states (see the analogous observation that was made for entanglement in Section~\ref{sec:product-frame}). The results presented here simply establish that in this specific setting these two inequivalent properties emerge at the same noise value. 

We further observe that, using results from Refs.~\cite{reichardt_quantum_2005, virmani_classical_2005} in a completely analogous way, one can also derive the inverse threshold for the T-gate under dephasing noise. This has the value $p_{\rm cl}=(\sqrt{2}-1)/2\sqrt{2}\approx 0.146$, which is compatible with the numerical results presented in Figure~\ref{fig:stabs_depol}.

\subsection{Diagonal product state frame}\label{app:analytics-prod}
In the continuous frame containing all product states, single-qubit unitaries are ``free'' in the sense that they map frame elements to other frame elements. Adding depolarizing noise to these gates maps frame elements to convex combinations of frame elements. It thus suffices to study the noisy $CNOT$ gate.
To bound the simulation cost associated with a noisy $CNOT$ gate, we aim to find optimal decompositions in terms of product states of the two-qubit density matrix 
\begin{equation}
\mathcal{N}_{p}^{\textrm{depol.} \, \otimes 2}\left(CNOT \, |\psi_{x}^{(1)}\rangle\langle\psi_{x}^{(1)}|\otimes|\psi_{x}^{(2)}\rangle\langle\psi_{x}^{(2)}|\, CNOT^\dag \right),
\end{equation} 
for arbitrary single qubit states $\psi_1$ and $\psi_2$ and local depolarizing noise $\mathcal{N}_p^{\textrm{depol.} \, \otimes 2}$.

Local unitaries on single qubits commute with local depolarizing noise and leave a decomposition's coefficient one-norm $||\lambda^{(CNOT,\rm{depol.})}||_1$ invariant. We may therefore assume without loss of generality that we have applied such local rotations to align the Schmidt vectors of the noise-free state to the computational basis. That is, it suffices to find optimal decompositions of
\begin{equation}\label{eq:mpsdecomp}
\mathcal{N}_{p}^{\textrm{depol.}\, \otimes2}\left(|\Psi\rangle\langle\Psi|\right) = \sum_{y\in\mathcal{X}}\lambda_{y}|\psi_y^{(1)}\rangle\langle \psi_y^{(1)}|\otimes|\psi_y^{(2)}\rangle\langle\psi_y^{(2)}|,
\end{equation}
where $|\Psi\rangle = \sqrt{\eta}|00\rangle+\sqrt{1-\eta}|11\rangle$. 

An optimal decomposition is one that minimizes
\begin{equation}\label{eq:mpsneg}
\|\lambda^{(CNOT,\rm{depol.})}\|_1 = \sum_{y\in\mathcal{X}}\left|\lambda_y\right|. 
\end{equation}
In the following, we compute the minimal value of $\|\lambda^{(CNOT,\rm{depol.})}\|_1$ in terms of the noise strength $p$ and Schmidt coefficient $\eta$. We do so by leveraging the duality established in Theorem~\ref{thm1}, first proving a lower bound on $\|\lambda^{(CNOT,\rm{depol.})}\|_1$, and subsequently finding an explicit decomposition that saturates this bound. 

\subsubsection*{Lower bound}
We derive a lower bound on $\|\lambda^{(CNOT,\rm{depol.})}\|_1$ by choosing the dual-feasible certificate
\begin{equation}
    W_{\textrm{diag.}}= 4|\Phi\rangle\langle\Phi| - \id^{\otimes 2},
\end{equation} 
where $|\Phi\rangle = (|00\rangle+|11\rangle)/\sqrt{2}$.
This choice can be intuitively motivated as follows.
Initially, consider as input the maximally entangled Bell state $\ket{\Psi}=\ket{\Phi}$. Since $\ket{\Phi}$ is invariant under the local symmetry $U\otimes \overline{U}$, an optimal dual-feasible $W$ remains optimal after applying the corresponding twirl
\begin{equation}
    W \rightarrow \int \mathrm{d}\mu_{\rm{Haar}}(U) (U\otimes \overline{U}) W (U \otimes \overline{U})^\dag,
\end{equation}
where $\mu_{\rm{Haar}}$ is the uniform Haar measure on the single-qubit unitary group.
It is well-known that the image of this twirling operator is two-dimensional, spanned by the operators $\id^{\otimes 2}, |\Phi\rangle\langle\Phi|$. As it turns out, the same $W_{\textrm{diag.}}$ gives a tight lower-bound for all $\eta \in [0,1]$. 

First, we check that $W_{\textrm{diag.}}$ is indeed a dual-feasible operator. This is the case since
\begin{align}
\label{eqn:prod_overlap}
    & |(\langle \psi^{(1)}| \otimes \langle \psi^{(2)} |) |\Phi \rangle |^2 \\[0.2em]
    = \, &  |(\langle 0^{(1)}| \otimes \langle \psi'^{(2)} |) |\Phi \rangle |^2 \\[0.2em]
    = \, & | \langle \psi'^{(2)} |0 \rangle|^2/2 \, \leq 1/2,
\end{align}
where we used the symmetry of $\ket{\Phi}$ in the first step.

Invoking Theorem~\ref{thm1}, we arrive at the following lower bound:
\begin{align}
    \|\lambda^{(CNOT,\rm{depol.})}\|_1 & \geq \tr(W_{\textrm{diag.}}\mathcal{N}_p^{\rm{depol.\, \otimes 2}}(|\Psi\rangle \langle \Psi |)) \\
    & = (1-4p)^2\tr(W_{\textrm{diag.}} |\Psi\rangle \langle \Psi |) \\
    & = (1-4p)^2(1 + 4\sqrt{\eta(1-\eta)}), \label{eq:prod_lower_bound}
\end{align}
where in the first step we used that the depolarizing channel is self-adjoint with respect to the Hilbert-Schmidt inner product and $\mathcal{N}_p^{\rm{depol.\, \otimes 2}}(W_{\textrm{diag.}})= (1-4p)^2W_{\textrm{diag.}}$.

We note that all noisy channels we consider have coefficient one-norm $\|\lambda\|_1\geq1$ (check $W = \id$). The analytic formula above is tight only up to the threshold, where the one-norm plateaus at 1. Also the analytic formulas discussed below should be understood as valid up to the threshold.

\subsubsection*{Upper bound}
We now construct an explicit decomposition that matches the lower bound just derived. 
Evaluating the effect of depolarizing noise in the Schmidt basis of $\ket{\Psi}$, we find

\begin{align}
\label{eqn:noisy_bell}
\mathcal{N}_{p}^{\rm{depol.} \, \otimes 2}(|\Psi\rangle\langle\Psi|)  
& \equiv 
\setlength{\arraycolsep}{8pt}
\begin{pmatrix} 
\rho_{00} & 0 & 0 & \gamma \\ 
0 & \delta & 0 & 0 \\ 
0 & 0 & \delta & 0 \\ 
\gamma & 0 & 0 & \rho_{11} 
\end{pmatrix},
\end{align}
where 
\begin{align}
\gamma & \coloneqq (1-4p)^2\sqrt{\eta(1-\eta)}, \\[0.2em]
\delta & \coloneqq 2p(1-2p), \\[0.2em]
\rho_{00} & \coloneqq (1-4p)\eta+ 4p^{2}, \\[0.2em]
\rho_{11} & \coloneqq (1-4p)(1-\eta)+4p^{2}.
\end{align}
For one, we observe that it is the off-diagonal components $\propto \gamma$ that in general make this operator non-separable. Additionally, the entry $|b_1\, b_2\rangle\langle b_3\, b_4|$ is non-vanishing iff $b_1 - b_2 = b_3 - b_4$. 

We look for an ensemble of product states that allows us to fix the problematic coherence terms $\propto \gamma$. 
To minimize the one-norm of the full decomposition, we intuitively want this ensemble to simultaneously populate all other non-zero entries of $\mathcal{N}_p^{\rm{depol.}\, \otimes 2}(|\Psi\rangle \langle \Psi|)$, while making sure that there is no overall contribution to the vanishing entries. 

A straightforward way to impose this is to consider single-qubit states of the form $|u_k\rangle \propto c_0 |0\rangle + c_1 e^{i\phi_k} |1\rangle$, with $\phi_k = k\pi/2$:
Then, the average
\begin{equation}
\label{eqn:prod_ensemble}
    \frac{1}{4} \sum_{k = 0}^3 |u_k\rangle\langle u_k |\otimes |\overline{u}_k\rangle\langle \overline{u}_k|
\end{equation}
has a phase factor 
\begin{equation}
\frac{1}{4}\sum_{k=0}^3 e^{ik\frac{\pi}{2}(b_1 - b_2 + b_4 - b_3)}
\end{equation}
in the component $|b_1\, b_2\rangle\langle b_3\, b_4|$, ensuring a non-vanishing contribution iff $b_1 - b_2 = b_3 - b_4$. We are free to choose the relative weights such that $|c_0|^2|c_1|^2 = \gamma$, in particular $c_0 = \eta^{1/4}$ and $c_1 = (1-\eta)^{1/4}$. Note that the terms in \eqref{eqn:prod_ensemble} are $+1$-eigenstates of $W_{\textrm{diag}}$.

For this choice of $c_0$, $c_1$, and accounting for the proper normalization factors, we find the decomposition
\begin{align}
\mathcal{N}_{p}^{\rm{depol.} \, \otimes 2} & (|\Psi\rangle\langle\Psi|)  \\ &= (1-4p)^{2}(\sqrt{\eta}+\sqrt{1-\eta})^{2}\, \cdot \nonumber \\ & \hspace{5em}\frac{1}{4}\sum_{k=0}^{3}|u_k\rangle\langle u_k |\otimes|\overline{u}_k\rangle\langle\overline{u}_k|  \nonumber \\ \nonumber
&\quad + \left[ \rho_{00} - (1-4p)^2\eta \right]|00\rangle\langle00| \\[0.4em]
&\quad + \left[\rho_{11} - (1-4p)^2(1-\eta)\right]|11\rangle\langle11|  \nonumber \\[0.4em]
\nonumber &\quad + \left[\delta - (1-4p)^2\sqrt{\eta(1-\eta)}\right]\, \cdot \\[0.4em]
\nonumber &\qquad \qquad \quad \left( |01\rangle \langle 01 |+ | 10 \rangle \langle 10|\right),
\end{align}
with one-norm matching the lower bound \eqref{eq:prod_lower_bound}. This can be easily verified by noting that only the last coefficient will be negative below the threshold.

\subsection{Dyadic product frame} \label{app:analytics_prod_dyad}
The diagonal product state frame is a strict subset of its dyadic variant. Therefore, for a fixed operator, the optimal decomposition in the dyadic product frame always has coefficient one-norm smaller or equal to that in the diagonal product state frame. However, this added redundancy comes at the cost of additional initial frame elements we must consider. A priori, some of these may require decompositions that void the benefit of including more frame elements. 

In this section, we study the above trade-off explicitly. 
First, we prove that by including non-diagonal product elements, we are in fact able to find decompositions of \eqref{eqn:noisy_bell} with smaller coefficient one-norm at noise values below $p_{\textrm{cl}} = (1-1/\sqrt{3})/4$. Like in Appendix~\ref{app:analytics-prod}, we will prove the optimality of these decompositions by supplying a matching dual certificate $W_{\textrm{dyad.}}$.
On top of that, we show that the new, non-Hermitian initial frame elements in fact do not require decompositions with a larger coefficient one-norm. While both the diagonal and dyadic product frames share the same threshold $p_{cl}$, our findings illustrate that including non-Hermitian frame elements offers a strict improvement below the threshold. This mirrors our results for stabilizer frames.

Following this roadmap, we begin by studying decompositions of the form
\begin{equation}
\label{eqn:dyad_decomp}
\mathcal{N}_{p}^{\textrm{depol.}\, \otimes2}\left(|\Psi\rangle\langle\Psi|\right) = \sum_{y\in\mathcal{X}}\lambda_y
|{\vphantom{\psi'}\psi}_y^{(1)}\rangle\langle{\psi'}_y^{(1)}|
\otimes
|{\vphantom{\psi'}\psi}_y^{(2)}\rangle\langle{\psi'}_y^{(2)}|,
\end{equation}
where $|\Psi\rangle = \sqrt{\eta}|00\rangle+\sqrt{1-\eta}|11\rangle$, without loss of generality. Recall that, in the Schmidt basis of $\Psi$, this operator is given by Equation~\eqref{eqn:noisy_bell}.

\subsubsection*{Lower bound}
Consider $W_{\textrm{dyad.}} = 2|\Phi\rangle\langle\Phi|$. By Inequality~\eqref{eqn:prod_overlap}, $W_{\textrm{dyad.}}$ is dual-feasible with respect to the dyadic product frame. The duality of Theorem~\ref{thm1} then implies
\begin{align}
\label{eqn:dyad_lowerbound}
    \|\lambda^{(CNOT)}\|_1 & \geq \tr(W_{\textrm{dyad.}}\mathcal{N}_p^{\textrm{depol.} \, \otimes 2}(|\Psi\rangle\langle\Psi|)) \\[0.2em]
    & = \dfrac{1}{2}\left[ 1+ (1-4p)^2(1 + 4\sqrt{\eta(1-\eta)}\,)\right], \\[-3em] \nonumber
\end{align}
where we used that 
\begin{align}
\mathcal{N}_p^{\textrm{depol.} \, \otimes 2}(W_{\textrm{dyad.}}) & = (1-4p)^2 W_{\textrm{dyad.}} \\ \nonumber & \qquad \qquad + \frac{1}{2}(1-(1-4p)^2) \, \id^{\otimes 2} \\[0.2em]
& = \frac{\id^{\otimes 2}}{2} + \frac{(1-4p)^2}{2} W_{\textrm{diag.}}.
\end{align}

\subsubsection*{Upper bound}
Next, we find explicit decompositions that saturate this lower bound. Unlike before, we can now directly populate the off-diagonal coherence terms $\propto \gamma$, as $|00\rangle\langle 11|$ and $|11\rangle\langle 00|$ are themselves frame elements. 

The idea is that we choose a suitable ensemble, similar to \eqref{eqn:prod_ensemble}, to first set the coherence terms to only $\kappa \coloneqq \min\{\gamma, \delta\}$. In that way, we are able to choose all coefficients to be positive, and we do not have to cancel unwanted contributions on penalty of a greater coefficient $1$-norm. This was not an option for the diagonal product state frame. Next, all incorrect entries can be fixed by adding a component proportional to the respective dyadic frame element to the decomposition. 

Defining $|v_k\rangle \coloneqq \sqrt{x}|0\rangle + \sqrt{1-x} e^{i\phi_k} |1\rangle$, $\phi_k = k\pi/2$, where we will specify $0 < x < 1$ shortly, we consider the decomposition
\begin{align}
\mathcal{N}_{p}^{\rm{depol.} \, \otimes 2} & (|\Psi\rangle\langle\Psi|)  \\ &= \frac{\kappa}{x(1-x)} \frac{1}{3}\sum_{k=0}^{3}|v_k\rangle\langle v_k |\otimes|\overline{v}_k\rangle\langle\overline{v}_k|  \nonumber \\ \nonumber
&\quad + \left(\rho_{00} - \kappa \, \frac{x}{1-x}\right)|00\rangle\langle00| \\[0.4em]
&\quad + \left(\rho_{11} - \kappa \, \frac{1-x}{x}\right)|11\rangle\langle11|  \nonumber \\[0.4em]
\nonumber &\quad + (\delta -\kappa) \left( |01\rangle \langle 01 |+ | 10 \rangle \langle 10|\right) \\[0.4em] \nonumber
&\quad + (\gamma - \kappa)\left( |00\rangle \langle 11 |+ | 11 \rangle \langle 00|\right).
\end{align} 
As can be readily verified, if we may choose $0<x<1$ such that all coefficients are positive, the coefficient $1$-norm of this decomposition matches the lower bound \eqref{eqn:dyad_lowerbound}, up to the threshold at $\delta = \gamma$.

Finally, we convince ourselves that a suitable choice of $x$ exists:
Demanding
\begin{align}
    \rho_{00} \geq \, \kappa \frac{x}{1-x}, \\[0.4em]
    \rho_{11} \geq \, \kappa \frac{1-x}{x},
\end{align}
we arrive at
\begin{align}
    \frac{\kappa}{\kappa + \rho_{11}} \leq x \leq \frac{\rho_{00}}{\kappa + \rho_{00}}.
\end{align}
This interval is in fact non-empty, since $\kappa^2 \leq \gamma^2 \leq \rho_{00}\rho_{11}$.

\subsubsection*{Non-Hermitian initial elements}
Finally, we prove that even for non-Hermitian initial frame elements, when we have to decompose $\mathcal{N}_{p}^{\textrm{depol.}\, \otimes2}\left(|\Psi\rangle\langle\Psi'|\right)$, the minimal coefficient one-norm never exceeds the maximum of \eqref{eqn:dyad_lowerbound}, achieved at $\eta = 1/2$. 

Let $\mathcal{F}$ denote the dyadic product frame and define $\|\cdot\|_{\mathcal{F}}$ like in Theorem~\ref{thm1}.
In the noiseless case, we know that 
\begin{align}
\label{eqn:noise_free_neg}
    \||\Psi\rangle\langle \Psi' |\|_{\mathcal{F}} & \leq (\sqrt{\eta} + \sqrt{1-\eta})(\sqrt{\eta'} + \sqrt{1-\eta'}) \\ \nonumber & \leq \sqrt{2}\, \sqrt{2} = 2,
\end{align}
which follows by considering the Schmidt decompositions of $\Psi$ and $\Psi'$.

Complementary to that, the single-qubit depolarizing channel $\mathcal{N}_p^{\textrm{depol.}}$ becomes entanglement breaking for a noise rate $p \geq 1/6$. This can be seen by noting that at $p = 1/6$, the channel is equivalent to the following measure-and-prepare channel:
\begin{align}
\label{eqn:prepmeas}
    \mathcal{N}_{p=1/6}^{\textrm{depol.}}(A)  = \frac{1}{3}\sum_{P \in \{X,Y,Z\}} \sum_{ \sigma \in \{\pm\}}&\Braket{\sigma_P |A| \sigma_P} \\[-0.5em] \nonumber & \qquad \cdot |\sigma_P\rangle\langle \sigma_P |,
\end{align}
where $|\sigma_P\rangle$ is the eigenstate of the Pauli operator $P$ with eigenvalue $\sigma$. That is, the channel is equivalent to measuring in a random Pauli eigenbasis and preparing the eigenstate corresponding to the measurement outcome, destroying all correlations with other subsystems. It therefore follows that 
\begin{align}
\label{eqn:free_depol}
    \|\mathcal{N}_{p=1/6}^{\textrm{depol.}}\otimes\mathcal{I}(|\Psi\rangle\langle \Psi' |)\|_{\mathcal{F}} \leq 1,
\end{align}
for any input dyad $|\Psi\rangle\langle \Psi' |$ as this will be mapped to a separable state.
Equality~\eqref{eqn:prepmeas} can be readily verified by noting that non-identity Pauli operators are indeed rescaled by $1/3$, since $\Braket{\sigma_P| Q |\sigma_P} = 0$ if $P\neq Q$.

Having established these edge cases, we note that we can decompose any depolarizing channel in terms of $\mathcal{I}$ and the entanglement breaking $\mathcal{N}_{p=1/6}^{\textrm{depol.}}$:
\begin{equation}
\label{eqn:depol_ch_decomp}
    \mathcal{N}_{q/4}^{\textrm{depol.}} = \left(1-6q\right)\mathcal{I} + 6q\,\mathcal{N}_{p=1/6}^{\textrm{depol.}}.
\end{equation}

Now, removing the fully depolarizing component from $\mathcal{N}_p^{\textrm{depol.} \, \otimes 2}$, and redistributing its weight, we consider the channel
\begin{equation}
    \mathcal{N}' = \frac{1}{2}\left(\mathcal{N}_{\textstyle \frac{1-(1-4p)^2}{4}}^{\mathrm{depol.}}\otimes \mathcal{I} + \mathcal{I} \otimes \mathcal{N}_{\textstyle \frac{1-(1-4p)^2}{4}}^{\mathrm{depol.}}\right),
\end{equation}
which agrees with our local noise model $\mathcal{N}_p^{\textrm{depol.}\, \otimes 2}$ on the Bell state $|\Phi\rangle\langle \Phi|$. In fact, the two differ only on input Paulis with weight $1$. On top of that, $\mathcal{N}'$ is a convex combination of single-qubit noise, and therefore simpler than our local depolarizing noise model.

Decomposing the single-qubit noise channels like \eqref{eqn:depol_ch_decomp}, and using our findings \eqref{eqn:noise_free_neg} and \eqref{eqn:free_depol}, we can bound
\begin{align}
\label{eqn:non_herm_bound}
    \|\mathcal{N}'(|\Psi\rangle\langle \Psi' |)\|_{\mathcal{F}} \leq 2\left(1-6q\right) + 6q,
\end{align}
with $q = (1-(1-4p)^2)/4$, where the right-hand side is exactly the lower bound~\eqref{eqn:dyad_lowerbound} for $\eta = 1/2$.

We now want to conclude that 
\begin{align}
    \|\mathcal{N}_p^{\textrm{depol.}\, \otimes 2}(|\Psi\rangle\langle \Psi' |)\|_{\mathcal{F}} & \leq \|\mathcal{N}'(|\Psi\rangle\langle \Psi' |)\|_{\mathcal{F}} \label{eq:ineq_N-N'}\\
    & \leq (1 + 3(1-4p)^2)/2.
\end{align}
For this, it remains to prove the first inequality~\eqref{eq:ineq_N-N'}. This can be shown by noting that $\mathcal{N}_p^{\textrm{depol.}\, \otimes 2}$ can be written as a sum
\begin{align}
    \mathcal{N}_p^{\textrm{depol.}\, \otimes 2} & = \alpha \mathcal{N}' + \beta \mathcal{F}\circ \mathcal{N}', \label{eq:decomp_N-alpha-beta}
\end{align}
where $\mathcal{F}$ is the ``free'' operation that negates weight-$1$ Paulis, while leaving all others invariant:
\begin{equation}
    \mathcal{F}(A) = (Y\otimes Y) A^T (Y\otimes Y).
\end{equation}
While $\mathcal{F}$ is not a quantum channel, it maps dyadic product frame elements to other frame elements. From this it follows in particular that 
\begin{align}
    \|\mathcal{F}\circ \mathcal{N}'(A)\|_{\mathcal{F}}\leq \|\mathcal{N}'(A)\|_{\mathcal{F}}.
\end{align}
The reason that the decomposition \eqref{eq:decomp_N-alpha-beta} is possible is exactly because both channels only differ on the weight-1 Pauli subspace.
Finally, the coefficients $\alpha, \beta \in \mathbb{R}$ are given by 
\begin{align}
    \alpha  & = \frac{1}{2}+ \frac{1-4p}{(1 + (1-4p)^2)}\\
    \beta & = \frac{1}{2}-\frac{1-4p}{(1 + (1-4p)^2)}.
\end{align}
These coefficients define a convex combination, from which the desired Inequality~\eqref{eq:ineq_N-N'} follows, completing the proof.

\subsection{Extended Pauli frame} \label{app:analytics-ext}
For $a\in[0,1]$, consider the Hermitian extended Pauli frame with local factors
\begin{equation}
  f_x^{(k)}
  \in \left\{\id,X,Y,Z,\;
    a\frac{X+Y}{\sqrt2},\;
    a\frac{X-Y}{\sqrt2}\right\}.
  \label{eq:frame}
\end{equation}

Further, consider a more general single-qubit noise channel $\mathcal{N}$, whose adjoint action in the Pauli basis is of the form
\begin{equation}
    \label{eq:gen_noise}
    \mathcal{N}^\dag \equiv 
    \begin{pmatrix}
        1 & & & v \\
        & \mu & & \\
        & & \mu & \\
        & & & u
    \end{pmatrix},
\end{equation}
where $|u| + |v| \leq 1$. In particular, depolarizing, dephasing, and amplitude damping noise are all of this type, with
\begin{alignat}{6}
    &\mu^{\rm (depol.)}  && = (1-4p) = \, && u^{\rm (depol.)} && , && \, \, v^{\rm (depol.)} && = 0,\, \\
    &\mu^{\rm (deph.)}  && = (1-2p), \, && u^{\rm (deph.)} && = 1 ,&& \, \,  v^{\rm (deph.)} && = 0, \\
    &\mu^{\rm(a.d.)} && = \sqrt{1-q} = \,   && \sqrt{ u^{\rm (a.d.)}} && , && \, \, v^{\rm (a.d.)} && = q.
\end{alignat}
The important feature that these noise channels have in common, and the one we will make use of in the following derivations, is that they are uniformly contractive in the Pauli $X,Y$-subspace, while also mapping Pauli $Z$ elements to a (sub-)convex combination. In fact, we can handle arbitrary right-most columns in \eqref{eq:gen_noise}, as long as their one-norm does not exceed 1.

We begin by calculating the cost of the noisy single-qubit $H$ and $T$ gates, making use of the duality derived in Appendix~\ref{sec:appendixA}:
Relabelling Paulis like $(P_0,P_1,P_2,P_3)=(\id,X,Y,Z)$, a single-qubit operator $W = \sum_{i=0}^3 \alpha_i P_i/\sqrt{2}$ is dual-feasible with respect to the extended Pauli frame iff
\begin{alignat}{2}
        & | \alpha_i | && \leq 1/\sqrt{2}, \label{eq:dual_feasibility_1}\\
        & | \alpha_1 | +  |\alpha_2 | \label{eq:dual_feasibility_2} && \leq 1/a.
\end{alignat}
Here, Inequality~\eqref{eq:dual_feasibility_1} bounds the overlap with regular Pauli operators, whereas Inequality~\eqref{eq:dual_feasibility_2} bounds the overlap with the rotated elements of the extension.
We denote the set of Hermitian single-qubit operators $W$ satisfying Inequalities~\eqref{eq:dual_feasibility_1} and \eqref{eq:dual_feasibility_2} with $\mathcal{W}_1$. The dual-feasible set $\mathcal{W}_1$ can be parametrized by the four-component vector $\boldsymbol{\alpha}$. 

For the noisy $T$-gate, since $T^2$ is a Clifford unitary, it suffices to consider the initial frame elements $X,Y$. In the Heisenberg picture, the noisy version maps these to
\begin{equation}
\label{eq:backpropagation_T}
    X, Y \mapsto  \frac{\mu}{\sqrt{2}}(X \pm Y).
\end{equation}
Then, by Theorem~\ref{thm1}, the optimal decomposition with respect to the extended Pauli frame satisfies
\begin{align}
 \|\lambda^{(T)}\|_1  & = \mu \cdot \max_{\boldsymbol{\alpha}\in \mathcal{W}_1}|\alpha_1 \pm \alpha_2 | \\[0.2em]
 & = \mu \cdot \max_{\boldsymbol{\alpha}\in \mathcal{W}_1}|\alpha_1| + | \alpha_2| \label{eq:gen_Pauli_T_1} \\
   & = \mu \cdot \min\{\sqrt{2}, 1/a\} \label{eq:gen_Pauli_T_2}.
\end{align}
The minimum appears because the sum in Equation~\eqref{eq:gen_Pauli_T_1} can be as large as $1/a$, but only if Condition~\eqref{eq:dual_feasibility_1} remains satisfied. If $1/a \geq 2/\sqrt{2}$, the maximum value levels off at $\sqrt{2}$. 
In these two cases, depending on $a$, an optimal decomposition is either the regular Pauli expansion, or one with a single element proportional to \eqref{eq:backpropagation_T}, respectively. 

Since the adjoint of any trace-preserving channel is unital, we note that all noisy channels we consider have coefficient one-norm $\|\lambda\|_1\geq1$. For simplicity, in order to avoid nested $\max$ and $\min$, we present analytic formulas only up to the threshold, where $\|\lambda\|_1$ plateaus at 1.

For the $H$-gate, itself Clifford, we must only consider initial frame elements coming from the extension of the Pauli basis. An analogous analysis yields 
\begin{align}
 \|\lambda^{(H)}\|_1  & = a  \mu \cdot \max_{\boldsymbol{\alpha}\in \mathcal{W}_1} |\mp \alpha_2 + \alpha_3 |\\[0.2em]
 & = a \mu \cdot \max_{\boldsymbol{\alpha}\in \mathcal{W}_1} |\alpha_2| + |\alpha_3| \\[0.2em]
   & = a\mu \cdot (\min\{1/a, 1/\sqrt{2}\} + 1/\sqrt{2}). 
   \label{eq:gen_Pauli_H}
\end{align}
To get the correct noise factor, note that in our convention noise acts after the ideal unitary, with that order reversed in the Heisenberg picture.
An explicit decomposition achieving this value, depending on $a$, is either the linear combination of Pauli $Z$ and both elements in the extension, or the regular Pauli expansion, respectively.

As can be read-off from Equations~\eqref{eq:gen_Pauli_T_2} and \eqref{eq:gen_Pauli_H}, increasing $a$ reduces the cost of simulating noisy $T$ gates, but has the opposite effect for noisy $H$ gates. Setting $a = 2^{-1/4}$ balances the two, and yields
\begin{equation}
    \|\lambda^{(T)}\|_1 = \|\lambda^{(H)}\|_1 = 2^{1/4}\mu.
\end{equation}

It remains to study the noisy two-qubit $CNOT$ gate, also itself Clifford.
We must therefore examine three types of initial frame elements, now on two qubits: 
\begin{enumerate}
    \item ``Extended'' Pauli on control, Pauli on target: $$a\frac{X\pm Y}{\sqrt{2}}\otimes P,$$
    \item ``extended'' Pauli on control and target: $$a\frac{X \pm Y}{\sqrt{2}}\otimes a\frac{X \pm Y}{\sqrt{2}},$$
    \item Pauli on control, ``extended'' Pauli on target: $$P \otimes a\frac{X+Y}{\sqrt{2}}.$$
\end{enumerate}
The first type of initial frame elements is unproblematic: under back-propagation with a $CNOT$ gate, these elements are mapped to other frame elements. To see this, recall that $CNOT$ propagates Pauli $X$ factors from control to target, and Pauli $Z$ factors from target to control. 

The second type can be dealt with in similar fashion:
we can rewrite these operators as a sum of two type $1$ terms, letting us bound the optimal coefficient one-norm $\|\lambda^{(CNOT)}\|_1$ coming from these initial frame elements by $2\mu^2 a/\sqrt{2}$. 

Finally, decomposing type 3 elements in the Pauli basis, each term is mapped to some signed Pauli under back-propagation, up to noise. We conclude
\begin{align}
   \|\lambda^{(CNOT)}\|_1 \leq \sqrt{2}a\mu.
\end{align}
Once we set $a=2^{-1/4}$ this matches the single-qubit expressions $\|\lambda^{(T)}\|_1,\|\lambda^{(H)}\|_1$.

\subsubsection*{Optimality}
Not only does the extended Pauli frame improve over back-propagation in the Pauli basis, but it is in fact the optimal choice among all normalized product frames. By normalized product frame we mean a frame of the form~\eqref{eq:generalized-pauli-frame}, where the local factors $f_x^{(j)}$ are normalized as $\|f_x^{(j)}\| \leq 1$ (note that this is a suitable normalization to guarantee that the Heisenberg picture condition~\eqref{eq:bounded-expectation-heis} is satisfied for all product initial states). We will indeed now show that any such normalized product frame gives
\begin{equation}
    \max\{\|\lambda^{(CNOT)}\|_1, \|\lambda^{(T)}\|_1\} \geq 2^{1/4}\mu,
\end{equation}
for all decompositions. 
The extended Pauli frame \eqref{eq:generalized-pauli-set} saturates this bound. 
To prove the bound, the dual characterization of Theorem~\ref{thm1} will again come in handy.

We begin by defining the two-qubit unitary $U_{ij} = CNOT_{j,i} T_i$, the qubit $j$ acting as control, and consider the quantum circuit on $k+1$ qubits defined by $U_{0,1} \dots U_{0,k-1} U_{0,k}$. This circuit will solely serve as a proxy to establish the desired lower bound. In particular, we make no claim of it being hard to simulate.

If we now take the individual $T$ and $CNOT$ gates to be noisy, characterized by a local noise model like \eqref{eq:gen_noise}. Denoting the noisy circuit's channel by $\mathcal{C}$, we for one find
\begin{align}
    \label{eq:gen_pauli_lower_bound}
    \tr(W \mathcal{C}^{*}(P)) &\leq \|\mathcal{C}^{*}(P) \|_\mathcal{F} \\
    &\leq \|P \|_\mathcal{F} \, \|\lambda^{(CNOT)}\|^k_1 \, \|\lambda^{(T)}\|^k_1,
\end{align}
where $P$ is some Pauli operator, and $W$ is a dual-feasible operator like in Theorem~\ref{thm1}. This inequality holds for all such $W$ and decompositions, and follows from the sub-multiplicativity of the norm $\|\cdot\|_\mathcal{F}$, defined in Theorem~\ref{thm1}.

Next, choosing $P = X \otimes \id \otimes \dots \otimes \id $, we construct $W$ to obtain an as tight as possible lower bound on the one-norms $\|\lambda^{(CNOT)}\|_1, \|\lambda^{(T)}\|_1$.
The back-evolved operator $\mathcal{C}^*(P)$ is
\begin{equation}
    \mathcal{C}^*(P) = \left(\frac{\mu^2}{\sqrt{2}}\right)^k \sum_{\boldsymbol{b}\in \{0,1\}^k} \sigma_{\boldsymbol{b}} Q_{b_k}\otimes Z^{b_1}\otimes \dots \otimes Z^{b_k},
\end{equation}
where $\sigma_{\boldsymbol{b}}\in\{\pm 1\}$, $Q_0 = X$, and $Q_1 = Y$. 
To maximize the left-hand side of Inequality~\eqref{eq:gen_pauli_lower_bound}, we define $W$ to include all the same terms as $\mathcal{C}^*(P)$, up to a constant factor ensuring dual-feasibility. In particular, we choose
\begin{equation}
    W \coloneqq \frac{1}{2^{k+1}} \sum_{\boldsymbol{b}\in\{0,1\}^k} \sigma_{\boldsymbol{b}} Q_{b_k} \otimes Z^{b_1} \otimes \dots \otimes Z^{b_k},
\end{equation}
and evaluate
\begin{equation}
    \tr(W\mathcal{C}^*(P))  = ( \sqrt{2}\mu^2)^k. 
\end{equation}
Therefore, if $W$ is dual-feasible, the following inequality is true for all $k$:
\begin{equation}
    \|P \|_\mathcal{F} \, \|\lambda^{(CNOT)}\|^k_1 \, \|\lambda^{(T)}\|^k_1 \geq (\sqrt{2}\mu^2)^k.
\end{equation}
Taking the $k$-th root and then the limit for $k\rightarrow +\infty$ of the last expression, it follows that 
\begin{equation}
    \|\lambda^{(CNOT)}\|_1 \, \|\lambda^{(T)}\|_1 \geq \sqrt{2}\mu^2,
\end{equation}
which can only be satisfied if 
\begin{equation}
    \max\{\|\lambda^{(CNOT)}\|_1, \|\lambda^{(T)}\|_1\} \geq 2^{1/4}\mu,
\end{equation}
as we intended to show. 

The only thing left is to verify that the Hilbert-Schmidt inner product of $W$ with any element of an arbitrary normalized product frame is bounded by 1. Let $F_x = \otimes_{j=0}^{k} f_{x}^{(j)}$ be a general such frame element. Then,
\begin{align}
    |\tr(WF_x)| & \leq \frac{1}{2^{k+1}} \Big|\sum_{\boldsymbol{b}\in\{0,1\}^k}\tr(Q_{b_k} f_x^{(0)})
    \\[-0.5em]  \nonumber 
    & \hspace{35mm} \prod_{j = 1}^k \tr(Z^{b_j}f_x^{(j)}) \Big| \\
    & \leq \Big| \sum_{b_k\in \{0,1\}} \frac{\tr(Q_{b_k} f_x^{(0)})}{2} \frac{\tr(Z^{b_k} f_x^{(k)})}{2} \Big| \\[-0.5em] \nonumber 
    & \hspace{35mm} \prod_{j = 1}^{k-1} \Big|\tr(\frac{\id+Z}{2}f_x^{(j)})\Big|\\[0.5em]
    &\leq  \sqrt{\sum_{b_k\in \{0,1\}} \Big|  \frac{\tr(Q_{b_k} f_x^{(0)})}{2} \Big|^2 } \\[-0.5em] \nonumber 
    & \hspace{20mm} \sqrt{\sum_{b_k'\in \{0,1\}} \Big|   \frac{\tr(Z^{b_k'} f_x^{(k)})}{2} \Big|^2 } \\[-0.5em] \nonumber 
    & \hspace{35mm} \prod_{j = 1}^{k-1} \left|\braket{0|f_x^{(j)}|0})\right|\\
    & \leq \frac{1}{2} \|f_x^{(0)} \|_{2} \|f_x^{(k)}\|_2 \prod_{j = 1}^{k-1} \|f_x^{(j)}\|\\
    &\leq 1,
\end{align}
where we have first used the Cauchy-Schwarz inequality on the two-dimensional vectors $\left(\tr(Q_{b_k}f_x^{(0)})/2\right)_{b_k}$ and $\left(\tr(Z^{b_k}f_x^{(k)})/2\right)_{b_k}$, and then $\|f\|_2^2=\tr(f^\dagger f)=2\sum_{i=0}^3|\tr(P_{i}f)/2|^2$. Finally, we have also used $\|f\|_2\leq \sqrt{2}\|f\|$ for single qubit operators.

\section{No-go theorems from magic distillation}\label{app:nogo}

When analysing stabilizer frames in Section~\ref{sec:stabiliser-frames}, we observed a simulability threshold $p_{\rm cl} \approx 0.11$ that was apparently independent of the frame choice (whether diagonal or dyadic). This fact can be explained by a observation first made by Reichardt in Ref.~\cite{reichardt_quantum_2009}. There, he showed that the following statement is true. Any algorithm that is able to simulate circuits composed of noiseless adaptive Clifford gates and noisy $T$-gates, with noise rate below a certain threshold, is also able to simulate universal quantum computation.

In this context, the term \emph{adaptive Clifford gates} refers to $q$-qubit unitary Clifford gates which are applied conditioned on the classical outcome of a Pauli measurement on another qubit. One can collectively see this operation as a non-unitary quantum channel $\cC$ applied to $q+1$ qubits. It is not hard to see that our stabiliser frame algorithms are able to efficiently simulate these noiseless adaptive Clifford channels, similarly to the case of regular Clifford gates. For what concerns the noise threshold mentioned above, this is referred to as the \emph{magic distillation threshold} and for depolarised $T$-gates is known to be $p_{\rm dist}=(6-2\sqrt{2})/28\approx 0.11$~\cite{reichardt_quantum_2005,reichardt_quantum_2009}. We conclude that, if the stabiliser frame algorithms were able to efficiently simulate the noisy $T$-gate at noise strengths lower than $p_{\rm dist}$, then they would also be able to efficiently classically simulate universal quantum computation, which we do not believe to be possible. In this sense, the thresholds $p_{\rm cl} =(6-2\sqrt{2})/28$ observed for the the stabiliser frame methods are optimal.

The statement of Reichardt above is proven by observing that circuits composed of noiseless adaptive Clifford gates and noisy $T$-gates below the distillation threshold can be used to efficiently simulate arbitrary universal quantum computations~\cite{BK-universalQC}. Indeed, if one has access to noisy $T$-gates, one can use them to create noisy approximations of the $T$-state vector $\ket{T}=T\ket{+}$. The fact that the noise rate is below the distillation threshold then implies that a logarithmically large number of these noisy $T$-state can be combined, using a circuit of noiseless adaptive Clifford gates, to produce (or \emph{distil}) an arbitrarily good approximation of the ideal $T$-state. This $T$-state can then be used to perform magic state injection (again by using a noiseless adaptive Clifford circuit) and reproduce the action of the noiseless $T$-gate, which is the only missing element to achieve universality with the considered gate-set.

One may wonder if an analogous result can be derived from the concept of \emph{entanglement distillation}, which would be applicable to the product state frame discussed in Section~\ref{sec:product-frame}. Consider, indeed, a model where each circuit qubit is interpreted as a local party, who is allowed to perform arbitrary adaptive local gates. One can verify that these adaptive local gates are efficiently simulable by a frame composed of product states. Suppose then that we supplement this model with the possibility of performing noisy entangling gates between the local parties. If the noise on these gates is below the entanglement distillation threshold, then the parties could use adaptive local operations to distil noiseless Bell pairs between them. These Bell states could then be used in a gate teleportation protocol to implement noiseless entangling gates and achieve universal quantum computation. This reasoning would imply that the product frame algorithms should not be able to efficiently simulate noisy entangling gates below the entanglement distillation threshold.

However, we encounter here a difficulty in this analysis. Indeed, in order for two local parties to be able to perform Bell state distillation, they need to be able hold at least two shared noisy Bell pairs. This means that each local party needs to have a local Hilbert space dimension large enough to hold the local halves of at least two Bell states, plus possible further ancillary degrees of freedom necessary for the gate teleportation protocol. However, in the model discussed above, the local parties only hold a single circuit qubit, which is clearly not enough for this. In conclusion, it seems that an analysis based on entanglement distillation can only constrain the power of product state frames if the local dimension of these states is large enough, while no bounds are known for the qubit case discussed in Section~\ref{sec:product-frame}. It is an open question whether bounds for qubit states can be derived in this way.

\bibliography{references}

\end{document}